# Variation Evolving for Optimal Control Computation, a Compact Way*

Sheng ZHANG, Jiang-Tao HUANG, Kai-Feng HE, and Fei LIAO

**Abstract** A compact version of the variation evolving method (VEM) is developed in the primal variable space for optimal control computation. Following the idea that originates from the Lyapunov continuous-time dynamics stability theory in the control field, the optimal solution is analogized to the stable equilibrium point of a dynamic system and obtained asymptotically through the variation motion. With the introduction of a virtual dimension, namely the variation time, the evolution partial differential equation (EPDE), which seeks the optimal solution with a theoretical guarantee, is developed for the optimal control problem (OCP) with free terminal states, and the equivalent optimality conditions with no employment of costates are established in the primal space. These conditions show that the optimal feedback control law is generally not analytically available because the optimal control is related to the future states. Since the derived EPDE is suitable to be computed with the semi-discrete method in the field of PDE numerical calculation, the optimal solution may be obtained by solving the resulting finite-dimensional initial-value problem (IVP).

## 1 Introduction

Optimal control is a discipline that studies the optimization on dynamic systems. It is closely related to the engineering and has been widely studied [1]. Because of their complexity, optimal control problems (OCPs) are usually solved numerically. The prevailing numerical methods may be divided into two types, namely, the direct methods and the indirect methods [2]. The direct methods discretize the control or/and state variables to obtain a nonlinear programming (NLP) problem [3-5]. These methods are easy to apply, whereas the results obtained are usually suboptimal [6]. The indirect methods transform the OCP to a boundary-value problem (BVP) through the optimality conditions [2, 7]. They may produce more precise results, but often suffer from significant numerical difficulty due to the ill-conditioning of the Hamiltonian dynamics; that is, the stability of the costate dynamics is adverse to that of the state dynamics [8], and this makes the computation difficult without a good initial guess [9].

The authors are with the Computational Aerodynamics Institution, China Aerodynamics Research and Development Center, Mianyang, 621000, China. (e-mail: zszhangshengzs@hotmail.com).

Practical effective ways to address the difficulty may be employing the evolutionary algorithms and metaheuristics [9], or combining the homotopy approach [10, 11]. Researchers, from a philosophical perspective of unity, try to uncover the connection between the direct and the indirect methods. Positive studies include the direct collocation method [6], the Runge-Kutta discretization method [12], and the pseudo-spectral (PS) method [13-15], which is popular in recent decades. As such methods blend the two types of methods in a dualization view, they inherit advantages of both types and blur their difference [16].

Theories in the dynamics and control field often enlighten strategies for the numerical optimization. The notion of differential flatness, arising from the non-linear control theory, is advanced for the optimal control computation since its proposal in Ref. [17]. With the non-linear coordinate transformation to reduce the variables, the OCPs are simplified and the solutions may be generated rapidly [18-20]. In the field of NLP, which is the static counterpart to the OCP, the dynamic methods based on the continuous-time ordinary differential equation (ODE) are prosperously developed [21, 22]. With these methods, NLP problems are transformed to the initial-value problems (IVPs) to be solved. Through the control parameterization technique, the OCP may be discretized and then solved by the dynamic method for NLP [23]. Interestingly, addressing the OCPs directly in a continuous dynamical manner will result in equations with the partial differential equation (PDE) form, which determines the evolution dynamics of variables. In Ref. [24], from the view of "differential distance", which might be more accurately called "variational distance", the evolution PDE (EPDE) for OCPs with Mayer performance index is proposed through the "method of gradient". In Ref. [25], a variation evolving method (VEM), inspired by the Lyapunov continuous-time dynamics stability theory, is developed for the optimal control computation, and the EPDE, upon a reconstructed unconstrained functional, is derived from the viewpoint of variation motion in typical OCPs. Using the well-known semi-discrete method for PDE numerical calculation [26], the finite-dimensional IVPs obtained may be solved with the mature ODE integration methods.

The VEM also synthesizes the direct and indirect methods, but from a new standpoint. Because the extremum of the OCP is theoretically guaranteed the equilibrium point of the deduced dynamic system, the optimal solution will be gradually approached. However, in the work of Ref. [25], besides the states and the controls, the costates are also introduced, which increases the complexity of the computation. In this paper, a compact version of the VEM that uses only the primal variables is developed and the corresponding EPDE is formulated. Both the VEM and the "method of gradient" in [24] take advantage of the variation motion for the solution of the OCPs. However, herein the more general Bolza performance index is considered, which avoids the introduction of extra variable when transforming to the Mayer form. Therefore, the EPDE, derived from a different approach in the following, is more common. In particular, the costate-free optimality conditions are established with the primal variables, and they characterize the fixed point of the PDE dynamic system. More importantly, under the frame of the infinite-dimensional Lyapunov principle, the theoretical convergence to the optimal solution is proved.

Throughout the paper, our work is built upon the assumption that the solution for the optimization problem exists. We do not describe the existing conditions for the purpose of brevity. Relevant studies may be found in [27, 28] and references therein. In the following, first preliminaries that state the infinite-dimensional Lyapunov theory and the principle of the VEM are presented. Then the compact VEM to solve the OCPs with free terminal states is developed. During this course, the costate-free optimality conditions are derived, and are proved equivalent to the classic conditions. After that, illustrative examples are solved to verify the effectiveness of the method. Besides, in-depth comments are presented before concluding the paper.

## 2 Preliminaries

### 2.1 Infinite-Dimensional Lyapunov Theory

The VEM is a newly developed method for optimal solutions. It is enlightened from the inverse consideration of the Lyapunov continuous-time dynamics stability theory in the control field [29]. As the theoretical foundation of this method, the infinite-dimensional Lyapunov principle is introduced as follows.

**Definition 2.1** Consider an infinite-dimensional dynamic system described by

$$\frac{\partial \boldsymbol{y}(x,t)}{\partial t} = \boldsymbol{f}(\boldsymbol{y}, \boldsymbol{p}, x)\,, \tag{1}$$

$$\frac{\mathrm{d}\boldsymbol{p}(t)}{\mathrm{d}t} = \boldsymbol{\Gamma}(\boldsymbol{y}, \boldsymbol{p})\,, \tag{2}$$

where $t$ is the time, $x \in \mathbb{R}$ is the independent variable, $\boldsymbol{y}(x) \in \mathbb{D}(x) \in \mathbb{R}^n(x)$ is the function vector of $x$, $\boldsymbol{p} \in \Omega \in \mathbb{R}^m$ is a state vector, $\boldsymbol{f}: \mathbb{D}(x) \times \Omega \times \mathbb{R} \to \mathbb{R}^n(x)$ and $\boldsymbol{\Gamma}: \mathbb{D}(x) \times \Omega \to \mathbb{R}^m$ are vector functionals, $\mathbb{D}(x)$ is a certain function set and $\Omega$ is a certain number set. If $(\hat{\boldsymbol{y}}(x), \hat{\boldsymbol{p}}) \in (\mathbb{D}(x) \times \Omega)$ satisfies $\boldsymbol{f}\left(\hat{\boldsymbol{y}}(x), \hat{\boldsymbol{p}}, x\right) = \boldsymbol{0}$ and $\boldsymbol{\Gamma}\left(\hat{\boldsymbol{y}}(x), \hat{\boldsymbol{p}}\right) = \boldsymbol{0}$, then $(\hat{\boldsymbol{y}}(x), \hat{\boldsymbol{p}})$ is called an equilibrium solution.

Note that Definition 2.1 allows $\boldsymbol{f}$ to be a functional, an extension of the function type used in Ref. [25]. Mathematically, an equilibrium solution of the dynamic system (1) and (2) is the fixed point of their right functionals.

**Definition 2.2** The equilibrium solution $(\hat{\boldsymbol{y}}(x), \hat{\boldsymbol{p}})$ is an asymptotically stable equilibrium solution in $(\mathbb{D}(x) \times \Omega)$ if for any initial conditions $\boldsymbol{y}(x,t)\big|_{t=0} = \tilde{\boldsymbol{y}}(x) \in \mathbb{D}(x)$ and $\boldsymbol{p}(t)\big|_{t=0} = \tilde{\boldsymbol{p}} \in \Omega$, there is $\lim\limits_{t \to +\infty} \left\| (\boldsymbol{y}(x,t), \boldsymbol{p}(t)) - (\hat{\boldsymbol{y}}(x), \hat{\boldsymbol{p}}) \right\|_\infty = 0$, where $\left\| (\boldsymbol{y}(x), \boldsymbol{p}) \right\|_\infty := \max\left( \sup\limits_x \left( \sum_{i=1}^n y_i^2 \right)^{1/2}, \left( \sum_{i=1}^m p_i^2 \right)^{1/2} \right)$ is the defined supremum norm.

With the same proof in Ref. [25], the following lemma may be established.

**Lemma 2.1** *For the infinite-dimensional dynamic system* (1) *and* (2)*, if there exists a continuously differentiable functional* $V:\mathbb{D}(x)\times\Omega\rightarrow\mathbb{R}$ *such that*

*i)* $V\left(\hat{\boldsymbol{y}}(x),\hat{\boldsymbol{p}}\right)=c$ *and* $V\left(\boldsymbol{y}(x),\boldsymbol{p}\right)>c$ *in* $(\mathbb{D}(x)\times\Omega)/\{(\hat{\boldsymbol{y}}(x),\hat{\boldsymbol{p}})\}$,

*ii)* $\dot{V}\left(\boldsymbol{y}(x),\boldsymbol{p}\right)\leq 0$ *in* $(\mathbb{D}(x)\times\Omega)$ *and* $\dot{V}\left(\boldsymbol{y}(x),\boldsymbol{p}\right)<0$ *in* $(\mathbb{D}(x)\times\Omega)/\{(\hat{\boldsymbol{y}}(x),\hat{\boldsymbol{p}})\}$,

*where* $c$ *is a constant, then* $(\boldsymbol{y}(x),\boldsymbol{p})=(\hat{\boldsymbol{y}}(x),\hat{\boldsymbol{p}})$ *is an asymptotically stable equilibrium solution in* $(\mathbb{D}(x)\times\Omega)$.

Lemma 2.1 is a generalization of the Lyapunov stability theorem, without requiring $V$ vanishing at the equilibrium. When employing Lemma 2.1 in the following, note that the variables are functions of $t$ (not $x$) and the time indicates the variation time $\tau$, a virtual dimension introduced to describe the variation evolution process.

### 2.2 Principle of VEM

The VEM analogizes the optimal solution of an OCP to the asymptotically stable equilibrium point of an infinite-dimensional dynamic system, and seeks such dynamics to minimize a specific performance index that acts as the Lyapunov functional. To implement the idea, the variation time $\tau$ is introduced to describe the process that a variable evolves to the optimal solution under the dynamics governed by the variation dynamic evolution equations, which may be presented in the form of the EPDE and the evolution differential equation (EDE) [25]. Figure 1 illustrates the variation evolution of the variable $x(t)$ in the VEM to solve the OCP. Through the variation motion, the initial guess of $x(t)$ will evolve to the optimal solution, and the optimality conditions will be gradually achieved.

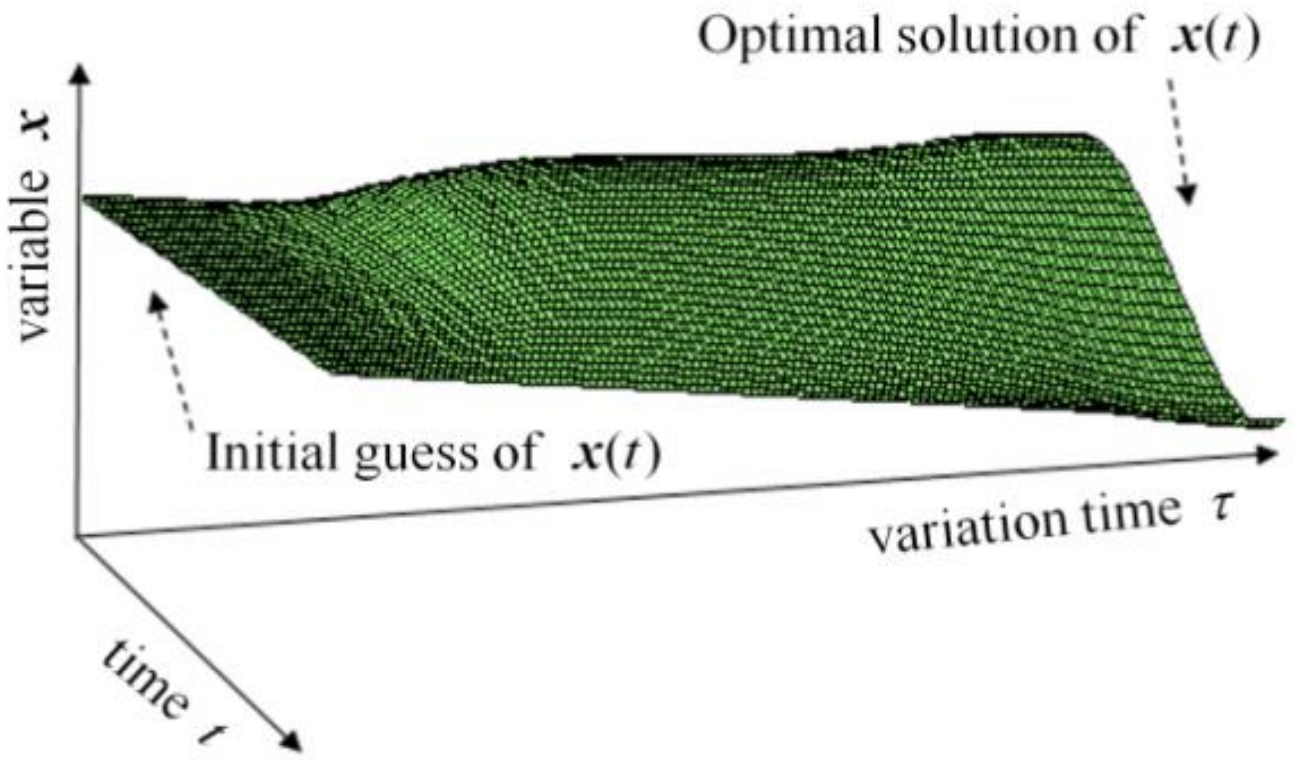

**Fig. 1** The illustration of the variable evolution along the variation time $\tau$ in the VEM [25]

For example, consider the following calculus-of-variations problem, which may be regarded as OCPs with integrator dynamics.

**Problem 1** For the following functional depending on the variable vector $\boldsymbol{y}(t) \in \mathbb{R}^n$

$$J = \int_{t_0}^{t_f} F\left(\boldsymbol{y}(t), \dot{\boldsymbol{y}}(t), t\right) \mathrm{d}t \,, \tag{3}$$

where $t \in \mathbb{R}$ is the time. The elements of $\boldsymbol{y}$ belong to $C^2[t_0, t_f]$, which denotes the set of variables with continuous second-order derivatives. The function $F : \mathbb{R}^n \times \mathbb{R}^n \times \mathbb{R} \to \mathbb{R}$ and its first-order and second-order partial derivatives are continuous with respect to $\boldsymbol{y}$, its time derivative $\dot{\boldsymbol{y}} = \dfrac{\mathrm{d}\boldsymbol{y}}{\mathrm{d}t}$ and $t$. $t_0$ and $t_f$ are the fixed initial and terminal times respectively, and the boundary conditions are prescribed as $\boldsymbol{y}(t_0) = \boldsymbol{y}_0$ and $\boldsymbol{y}(t_f) = \boldsymbol{y}_f$. Find the optimal solution $\hat{\boldsymbol{y}}$ that minimizes $J$, i.e.,

$$\hat{\boldsymbol{y}} = \arg\min(J) \,. \tag{4}$$

Following the idea of dynamic evolution to reduce some performance index, we anticipate that any initial guess of $\boldsymbol{y}(t)$, whose elements belong to $C^2[t_0, t_f]$, will evolve to the minimum solution through the variation motion. Like the decrease of the Lyapunov function for a stable dynamic system, if $J$ in Eq. (3) decreases with respect to the variation time $\tau$, i.e., $\dfrac{\delta J}{\delta \tau} \le 0$, we may eventually obtain the optimal solution. Differentiating Eq. (3) with respect to $\tau$ (even if $\tau$ does not explicitly exist) produces

$$\begin{aligned}\frac{\delta J}{\delta \tau} &= \int_{t_0}^{t_f} \left( {F_{\boldsymbol{y}}}^{\mathrm{T}} \frac{\delta \boldsymbol{y}}{\delta \tau} + {F_{\dot{\boldsymbol{y}}}}^{\mathrm{T}} \frac{\delta \dot{\boldsymbol{y}}}{\delta \tau} \right) \mathrm{d}t \\ &= {F_{\dot{\boldsymbol{y}}}}^{\mathrm{T}} \left. \frac{\delta \boldsymbol{y}}{\delta \tau} \right|_{t_f} - {F_{\dot{\boldsymbol{y}}}}^{\mathrm{T}} \left. \frac{\delta \boldsymbol{y}}{\delta \tau} \right|_{t_0} + \int_{t_0}^{t_f} \left( \left[ F_{\boldsymbol{y}} - \frac{\mathrm{d}}{\mathrm{d}t}(F_{\dot{\boldsymbol{y}}}) \right]^{\mathrm{T}} \frac{\delta \boldsymbol{y}}{\delta \tau} \right) \mathrm{d}t\end{aligned} \,, \tag{5}$$

where the column vectors $F_{\boldsymbol{y}} = \dfrac{\partial F}{\partial \boldsymbol{y}}$ and $F_{\dot{\boldsymbol{y}}} = \dfrac{\partial F}{\partial \dot{\boldsymbol{y}}}$ are the shorthand notations of partial derivatives. The superscript "$\mathrm{T}$" denotes the transpose operator. "$\left.\right|_{t_f}$" and "$\left.\right|_{t_0}$" mean "evaluated at $t_f$" and "evaluated at $t_0$", respectively. From an initial guess of $\boldsymbol{y}(t)$ that satisfies the boundary conditions at $t_0$ and $t_f$, then by enforcing $\dfrac{\delta J}{\delta \tau} \le 0$, we may set that

$$\frac{\delta \boldsymbol{y}}{\delta \tau} = -\boldsymbol{K} \left( F_{\boldsymbol{y}} - \frac{\mathrm{d}}{\mathrm{d}t}(F_{\dot{\boldsymbol{y}}}) \right), \; t \in ]t_0, t_f[ \,, \tag{6}$$

where $\boldsymbol{K}$ is an $n \times n$ dimensional positive-definite gain matrix. The variation dynamic evolution equation (6) may be considered from the view of the PDE formulation, by replacing the variation operation "$\delta$" and the differential operator "$\mathrm{d}$" with the partial differential operator "$\partial$" as

$$\frac{\partial \boldsymbol{y}(t,\tau)}{\partial \tau} = -\boldsymbol{K} \left( F_{\boldsymbol{y}} - \frac{\partial}{\partial t}(F_{\dot{\boldsymbol{y}}}) \right). \tag{7}$$

In addition, the initial condition is

$$\boldsymbol{y}(t,\tau)\big|_{\tau=0} = \tilde{\boldsymbol{y}}(t)\,, \tag{8}$$

where $\tilde{\boldsymbol{y}}(t) \in C^2[t_0, t_f]$ is an arbitrary solution that satisfies the boundary conditions.

**Theorem 2.1** *Solving the IVP defined by Eqs.* (7) *and* (8)*, when* $\tau \to +\infty$*,* $\boldsymbol{y}(t)$ *will satisfy the optimality conditions of Problem 1.*

***Proof*** For the infinite-dimensional dynamics governed by Eq. (7), Eq. (3) may act as the Lyapunov functional in $C^2[t_0, t_f]$. According to Lemma 2.1, the minimum solution of Problem 1 is an asymptotically stable equilibrium solution. Thus, when $\tau \to +\infty$, $\boldsymbol{y}(t)$ will meet the optimality condition, namely, the Euler-Lagrange equation [30, 31]

$$F_{\boldsymbol{y}} - \frac{\mathrm{d}}{\mathrm{d}t}(F_{\dot{\boldsymbol{y}}}) = \boldsymbol{0}\,. \tag{9}$$

QED. □

For the EPDE (7), its right function depends only on the time $t$ and is suitable to be solved with the semi-discrete method in the field of PDE numerical calculation. With the discretization along the normal time dimension, Eqs. (7) and (8) are transformed to an IVP with finite states; then the mature ODE numerical integration methods may be used to get the optimal solution. Note that the resulting IVP is defined with respect to the variation time $\tau$, not the normal time $t$. In the previous work [25], a demonstrative example is solved to verify the results.

## 3 The Compact VEM

### 3.1 Problem Definition

In this paper, we consider the following class of OCP that is defined as

**Problem 2** Consider the performance index of Bolza form

$$J = \varphi(\boldsymbol{x}(t_f), t_f) + \int_{t_0}^{t_f} L\big(\boldsymbol{x}(t), \boldsymbol{u}(t), t\big)\,\mathrm{d}t\,, \tag{10}$$

subject to the dynamic equation

$$\dot{\boldsymbol{x}} = \boldsymbol{f}(\boldsymbol{x}, \boldsymbol{u}, t)\,, \tag{11}$$

where $t \in \mathbb{R}$ is the time. $\boldsymbol{x} \in \mathbb{R}^n$ is the state vector, with its elements belonging to $C^2[t_0, t_f]$. $\boldsymbol{u} \in \mathbb{R}^m$ is the control vector, with its elements belonging to $C^1[t_0, t_f]$. The function $L: \mathbb{R}^n \times \mathbb{R}^m \times \mathbb{R} \to \mathbb{R}$ and its first-order partial derivatives are continuous with respect to $\boldsymbol{x}$, $\boldsymbol{u}$ and $t$. The function $\varphi: \mathbb{R}^m \times \mathbb{R} \to \mathbb{R}$ and its first-order and second-order partial derivatives are continuous with respect to $\boldsymbol{x}$ and $t$. The vector function $\boldsymbol{f}:\mathbb{R}^n\times\mathbb{R}^m\times\mathbb{R}\to\mathbb{R}^n$ and its first-order partial derivatives are continuous and Lipschitz in $\boldsymbol{x}$, $\boldsymbol{u}$ and $t$. The initial time $t_0$ is fixed and the terminal time $t_f$ is free. The initial boundary conditions are prescribed as

$$\boldsymbol{x}(t_0)=\boldsymbol{x}_0\,, \tag{12}$$

and the terminal states are free. Find the optimal solution $(\hat{\boldsymbol{x}},\hat{\boldsymbol{u}})$ that minimizes $J$, i.e.,

$$(\hat{\boldsymbol{x}},\hat{\boldsymbol{u}})=\arg\min(J)\,. \tag{13}$$

### 3.2 Derivation of EPDE

In Ref. [25], the problem is addressed by constructing an unconstrained functional problem that has the same extremum. This operation may be practical but it introduces the costate vector, which has the same dimension as the state vector. To avoid the resulting complexity, here we will address Problem 2 directly in the primal space, as we treated Problem 1 in the preceding. Consider the problem within the feasible solution domain $\mathbb{D}_o$, in which any solution satisfies Eqs. (11) and (12). First, we transform the general Bolza performance index (10) to the equivalent Lagrange type, i.e.,

$$\begin{aligned}J&=\varphi(\boldsymbol{x}(t_f),t_f)+\int_{t_0}^{t_f}L\big(\boldsymbol{x}(t),\boldsymbol{u}(t),t\big)\mathrm{d}t\\&=\varphi(\boldsymbol{x}(t_0),t_0)+\int_{t_0}^{t_f}\big(\varphi_t+\varphi_{\boldsymbol{x}}{}^{\mathrm{T}}\boldsymbol{f}(\boldsymbol{x},\boldsymbol{u},t)\big)\mathrm{d}t+\int_{t_0}^{t_f}L\big(\boldsymbol{x}(t),\boldsymbol{u}(t),t\big)\mathrm{d}t\,,\\&=\varphi(\boldsymbol{x}(t_0),t_0)+\int_{t_0}^{t_f}\big(\varphi_t+\varphi_{\boldsymbol{x}}{}^{\mathrm{T}}\boldsymbol{f}(\boldsymbol{x},\boldsymbol{u},t)+L(\boldsymbol{x},\boldsymbol{u},t)\big)\mathrm{d}t\end{aligned} \tag{14}$$

where $\varphi_t$ and $\varphi_{\boldsymbol{x}}$ are the partial derivatives. Note that the term $\varphi(\boldsymbol{x}(t_0),t_0)$ is constant and may be neglected in the performance index. Similarly, we differentiate Eq. (14) with respect to the variation time $\tau$ to obtain

$$\frac{\delta J}{\delta\tau}=(\varphi_t+\varphi_{\boldsymbol{x}}{}^{\mathrm{T}}\boldsymbol{f}+L)\Big|_{t_f}\frac{\delta t_f}{\delta\tau}+\int_{t_0}^{t_f}\left((\varphi_{t\boldsymbol{x}}{}^{\mathrm{T}}+\boldsymbol{f}^{\mathrm{T}}\varphi_{\boldsymbol{xx}}+\varphi_{\boldsymbol{x}}{}^{\mathrm{T}}\boldsymbol{f}_{\boldsymbol{x}}+L_{\boldsymbol{x}}{}^{\mathrm{T}})\frac{\delta\boldsymbol{x}}{\delta\tau}+(\varphi_{\boldsymbol{x}}{}^{\mathrm{T}}\boldsymbol{f}_{\boldsymbol{u}}+L_{\boldsymbol{u}}{}^{\mathrm{T}})\frac{\delta\boldsymbol{u}}{\delta\tau}\right)\mathrm{d}t\,, \tag{15}$$

where $\varphi_{t\boldsymbol{x}}$ and $\varphi_{\boldsymbol{xx}}$ are the second-order partial derivatives in the form of (column) vector and matrix respectively, and $\boldsymbol{f}_{\boldsymbol{x}}$ and $\boldsymbol{f}_{\boldsymbol{u}}$ are the Jacobian matrixes. Different from the calculus-of-variations problems, $\frac{\delta\boldsymbol{x}}{\delta\tau}$ and $\frac{\delta\boldsymbol{u}}{\delta\tau}$ are related because the profiles of $\boldsymbol{x}$ are determined by $\boldsymbol{u}$. For the solutions in $\mathbb{D}_o$, they need to satisfy the following variation equation

$$\frac{\delta\dot{\boldsymbol{x}}}{\delta\tau}=\boldsymbol{f}_{\boldsymbol{x}}\frac{\delta\boldsymbol{x}}{\delta\tau}+\boldsymbol{f}_{\boldsymbol{u}}\frac{\delta\boldsymbol{u}}{\delta\tau}\,, \tag{16}$$

with the initial condition $\left.\frac{\delta\boldsymbol{x}}{\delta\tau}\right|_{t_0}=\boldsymbol{0}$. Note that $\boldsymbol{f}_{\boldsymbol{x}}$ and $\boldsymbol{f}_{\boldsymbol{u}}$ are time-dependent matrixes linearized at the feasible solution $\boldsymbol{x}(t)$ and $\boldsymbol{u}(t)$. Eq. (16) is linear with respect to the variables $\frac{\delta\boldsymbol{x}}{\delta\tau}$ and $\frac{\delta\boldsymbol{u}}{\delta\tau}$, and has a zero initial value. Thus, it satisfies the superposition principle, and its solution may be explicitly expressed.

**Lemma 3.1** [32] *For the linear time-varying system*

$$\dot{\boldsymbol{x}} = \boldsymbol{A}(t)\boldsymbol{x} + \boldsymbol{B}(t)\boldsymbol{u} \,, \tag{17}$$

*with the initial value* $\boldsymbol{x}(t_0) = \boldsymbol{0}$, *where* $\boldsymbol{x} \in \mathbb{R}^n$ *is the state vector,* $\boldsymbol{u} \in \mathbb{R}^m$ *is the control vector, and* $\boldsymbol{A}(t)$ *and* $\boldsymbol{B}(t)$ *are the right dimensional coefficient matrixes, the solution is*

$$\boldsymbol{x}(t) = \int_{t_0}^{t} \boldsymbol{H}(t,s)\boldsymbol{u}(s)\,\mathrm{d}\,s \,, \tag{18}$$

*where* $\boldsymbol{H}(t,s)$ *is the* $n \times m$ *dimensional impulse response function that satisfies*

$$\boldsymbol{H}(t,s) = \begin{cases} \boldsymbol{\Phi}(t,s)\boldsymbol{B}(s) & t \geq s \\ \boldsymbol{0}_{n\times m} & s < t \end{cases} , \tag{19}$$

*and* $\boldsymbol{\Phi}(t,s)$ *is the* $n \times n$ *dimensional state transition matrix for the system from time point* $s$ *to time point* $t$, *which satisfies*

$$\frac{\partial \boldsymbol{\Phi}(t,s)}{\partial t} = \boldsymbol{A}(t)\boldsymbol{\Phi}(t,s) \,, \tag{20}$$

$$\frac{\partial \boldsymbol{\Phi}(t,s)}{\partial s} = -\boldsymbol{\Phi}(t,s)\boldsymbol{A}(s) \,, \tag{21}$$

$$\boldsymbol{\Phi}(t,t) = \boldsymbol{I}_{n\times n} \,, \tag{22}$$

*where* $\boldsymbol{I}_{n\times n}$ *is the* $n \times n$ *dimensional identity matrix.*

According to Lemma 3.1, Eq. (16) has the solution

$$\frac{\delta \boldsymbol{x}}{\delta \tau} = \int_{t_0}^{t} \boldsymbol{H}_o(t,s) \frac{\delta \boldsymbol{u}}{\delta \tau}(s)\,\mathrm{d}\,s \,, \tag{23}$$

where $\boldsymbol{H}_o(t,s)$ is the impulse response function corresponding to the specific $\boldsymbol{f}_{\boldsymbol{x}}(t)$ and $\boldsymbol{f}_{\boldsymbol{u}}(t)$. Substitute Eq. (23) into Eq. (15); there is

$$\begin{aligned} \frac{\delta J}{\delta \tau} &= (\varphi_t + \varphi_{\boldsymbol{x}}{}^{\mathrm{T}} \boldsymbol{f} + L)\Big|_{t_f} \frac{\delta t_f}{\delta \tau} + \int_{t_0}^{t_f} \left\{ \left( \varphi_{t\boldsymbol{x}}{}^{\mathrm{T}} + \boldsymbol{f}^{\mathrm{T}} \varphi_{\boldsymbol{xx}} + \varphi_{\boldsymbol{x}}{}^{\mathrm{T}} \boldsymbol{f}_{\boldsymbol{x}} + L_{\boldsymbol{x}}{}^{\mathrm{T}} \right) \left( \int_{t_0}^{t} \boldsymbol{H}_o(t,s) \frac{\delta \boldsymbol{u}}{\delta \tau}(s)\,\mathrm{d}\,s \right) + (\phi_{\boldsymbol{x}}{}^{\mathrm{T}} \boldsymbol{f}_u + L_{\boldsymbol{u}}{}^{\mathrm{T}}) \frac{\delta \boldsymbol{u}}{\delta \tau} \right\} \mathrm{d}t \\ &= (\varphi_t + \varphi_{\boldsymbol{x}}{}^{\mathrm{T}} \boldsymbol{f} + L)\Big|_{t_f} \frac{\delta t_f}{\delta \tau} + \int_{t_0}^{t_f} \int_{t_0}^{t} \left\{ \left( \varphi_{t\boldsymbol{x}}{}^{\mathrm{T}} + \boldsymbol{f}^{\mathrm{T}} \varphi_{\boldsymbol{xx}} + \varphi_{\boldsymbol{x}}{}^{\mathrm{T}} \boldsymbol{f}_{\boldsymbol{x}} + L_{\boldsymbol{x}}{}^{\mathrm{T}} \right) \boldsymbol{H}_o(t,s) \frac{\delta \boldsymbol{u}}{\delta \tau}(s) \right\} \mathrm{d}\,s\,\mathrm{d}\,t + \int_{t_0}^{t_f} \left( (\varphi_{\boldsymbol{x}}{}^{\mathrm{T}} \boldsymbol{f}_u + L_{\boldsymbol{u}}{}^{\mathrm{T}}) \frac{\delta \boldsymbol{u}}{\delta \tau} \right) \mathrm{d}t \end{aligned} . \tag{24}$$

By exchanging the order in the double integral, we may derive the following transformation as

$$\begin{aligned} &\int_{t_0}^{t_f} \int_{t_0}^{t} \left\{ \left( \varphi_{t\boldsymbol{x}}{}^{\mathrm{T}}(t) + \boldsymbol{f}^{\mathrm{T}}(t) \varphi_{\boldsymbol{xx}}(t) + \varphi_{\boldsymbol{x}}{}^{\mathrm{T}}(t) \boldsymbol{f}_{\boldsymbol{x}}(t) + L_{\boldsymbol{x}}{}^{\mathrm{T}}(t) \right) \boldsymbol{H}_o(t,s) \frac{\delta \boldsymbol{u}}{\delta \tau}(s) \right\} \mathrm{d}\,s\,\mathrm{d}\,t \\ &= \int_{t_0}^{t_f} \left( \int_{s}^{t_f} \left( \varphi_{t\boldsymbol{x}}{}^{\mathrm{T}}(t) + \boldsymbol{f}^{\mathrm{T}}(t) \varphi_{\boldsymbol{xx}}(t) + \varphi_{\boldsymbol{x}}{}^{\mathrm{T}}(t) \boldsymbol{f}_{\boldsymbol{x}}(t) + L_{\boldsymbol{x}}{}^{\mathrm{T}}(t) \right) \boldsymbol{H}_o(t,s)\,\mathrm{d}\,t \right) \frac{\delta \boldsymbol{u}}{\delta \tau}(s)\,\mathrm{d}\,s \end{aligned} . \tag{25}$$

To make the result clearer, the variable symbols are changed as $t \to \sigma$ and $s \to t$, without altering the final result of Eq. (25), to produce

$$\begin{aligned}&\int_{t_0}^{t_f}\left(\int_s^{t_f}\left(\varphi_{tx}^{\mathrm{T}}(t)+\boldsymbol{f}^{\mathrm{T}}(t)\varphi_{\boldsymbol{xx}}(t)+\varphi_{\boldsymbol{x}}^{\mathrm{T}}(t)\boldsymbol{f}_{\boldsymbol{x}}(t)+L_{\boldsymbol{x}}^{\mathrm{T}}(t)\right)\boldsymbol{H}_o(t,s)\,\mathrm{d}t\right)\frac{\delta\boldsymbol{u}}{\delta\tau}(s)\,\mathrm{d}s\\ &=\int_{t_0}^{t_f}\left(\int_t^{t_f}\left(\varphi_{tx}^{\mathrm{T}}(\sigma)+\boldsymbol{f}^{\mathrm{T}}(\sigma)\varphi_{\boldsymbol{xx}}(\sigma)+\varphi_{\boldsymbol{x}}^{\mathrm{T}}(\sigma)\boldsymbol{f}_{\boldsymbol{x}}(\sigma)+L_{\boldsymbol{x}}^{\mathrm{T}}(\sigma)\right)\boldsymbol{H}_o(\sigma,t)\,\mathrm{d}\sigma\right)\frac{\delta\boldsymbol{u}}{\delta\tau}(t)\,\mathrm{d}t\end{aligned}. \tag{26}$$

Thus, Eq. (24) may be reformulated as

$$\frac{\delta J}{\delta\tau}=(\varphi_t+\varphi_{\boldsymbol{x}}^{\mathrm{T}}\boldsymbol{f}+L)\Big|_{t_f}\frac{\delta t_f}{\delta\tau}+\int_{t_0}^{t_f}\left\{\left(\int_t^{t_f}\left(\varphi_{tx}^{\mathrm{T}}(\sigma)+\boldsymbol{f}^{\mathrm{T}}(\sigma)\varphi_{\boldsymbol{xx}}(\sigma)+\varphi_{\boldsymbol{x}}^{\mathrm{T}}(\sigma)\boldsymbol{f}_{\boldsymbol{x}}(\sigma)+L_{\boldsymbol{x}}^{\mathrm{T}}(\sigma)\right)\boldsymbol{H}_o(\sigma,t)\,\mathrm{d}\sigma\right)+(\varphi_{\boldsymbol{x}}^{\mathrm{T}}\boldsymbol{f}_u+L_{\boldsymbol{u}}^{\mathrm{T}})\right\}\frac{\delta\boldsymbol{u}}{\delta\tau}(t)\,\mathrm{d}t\,. \tag{27}$$

Now to achieve $\frac{\delta J}{\delta\tau}\le 0$, we may derive the variation dynamic evolution equation for the control $\boldsymbol{u}$ as

$$\frac{\delta\boldsymbol{u}}{\delta\tau}=-\boldsymbol{K}\left\{L_{\boldsymbol{u}}+\boldsymbol{f}_u^{\mathrm{T}}\varphi_{\boldsymbol{x}}+\left(\int_t^{t_f}\boldsymbol{H}_o^{\mathrm{T}}(\sigma,t)\left(L_{\boldsymbol{x}}(\sigma)+\varphi_{tx}(\sigma)+\varphi_{\boldsymbol{xx}}^{\mathrm{T}}(\sigma)\boldsymbol{f}(\sigma)+\boldsymbol{f}_{\boldsymbol{x}}^{\mathrm{T}}(\sigma)\varphi_{\boldsymbol{x}}(\sigma)\right)\mathrm{d}\sigma\right)\right\}, \tag{28}$$

and for the terminal time $t_f$ as

$$\frac{\delta t_f}{\delta\tau}=-k_{t_f}\left(L+\varphi_t+\varphi_{\boldsymbol{x}}^{\mathrm{T}}\boldsymbol{f}\right)\Big|_{t_f}, \tag{29}$$

where $\boldsymbol{K}$ is an $m\times m$ dimensional positive-definite gain matrix and $k_{t_f}$ is a scalar positive gain parameter.

**Remark 3.1** With the property of the state transition matrix, it may be obtained that

$$\begin{aligned}&\int_t^{t_f}\boldsymbol{H}_o^{\mathrm{T}}(\sigma,t)\left(\varphi_{tx}(\sigma)+\varphi_{\boldsymbol{xx}}^{\mathrm{T}}(\sigma)\boldsymbol{f}(\sigma)+\boldsymbol{f}_{\boldsymbol{x}}(\sigma)^{\mathrm{T}}\varphi_{\boldsymbol{x}}(\sigma)\right)\mathrm{d}\sigma\\ &=\int_t^{t_f}\frac{\mathrm{d}}{\mathrm{d}\sigma}\left(\boldsymbol{H}_o^{\mathrm{T}}(\sigma,t)\varphi_{\boldsymbol{x}}(\sigma)\right)\mathrm{d}\sigma\\ &=\boldsymbol{H}_o^{\mathrm{T}}(t_f,t)\varphi_{\boldsymbol{x}}(t_f)-\boldsymbol{f}_u^{\mathrm{T}}(t)\varphi_{\boldsymbol{x}}(t)\end{aligned}. \tag{30}$$

Thus, Eqs. (27) and (28) may be equivalently reformulated as follows.

$$\frac{\delta J}{\delta\tau}=(\varphi_t+\varphi_{\boldsymbol{x}}^{\mathrm{T}}\boldsymbol{f}+L)\Big|_{t_f}\frac{\delta t_f}{\delta\tau}+\int_{t_0}^{t_f}\left\{\left(\int_t^{t_f}L_{\boldsymbol{x}}^{\mathrm{T}}(\sigma)\boldsymbol{H}_o(\sigma,t)\,\mathrm{d}\sigma\right)+\varphi_{\boldsymbol{x}}^{\mathrm{T}}(t_f)\boldsymbol{H}_o(t_f,t)+L_{\boldsymbol{u}}^{\mathrm{T}}\right\}\frac{\delta\boldsymbol{u}}{\delta\tau}(t)\,\mathrm{d}t\,, \tag{31}$$

$$\frac{\delta\boldsymbol{u}}{\delta\tau}=-\boldsymbol{K}\left\{L_{\boldsymbol{u}}+\boldsymbol{H}_o^{\mathrm{T}}(t_f,t)\varphi_{\boldsymbol{x}}(t_f)+\int_t^{t_f}\boldsymbol{H}_o^{\mathrm{T}}(\sigma,t)L_{\boldsymbol{x}}(\sigma)\,\mathrm{d}\sigma\right\}. \tag{32}$$

Similarly, using the partial differential operator "$\partial$" and the differential operator "$\mathrm{d}$" to reformulate the variation dynamic evolution equations (23), (28) and (29), we may obtain the following EPDE and EDE as

$$\frac{\partial}{\partial\tau}\begin{bmatrix}\boldsymbol{x}(t,\tau)\\ \boldsymbol{u}(t,\tau)\end{bmatrix}=\begin{bmatrix}\displaystyle\int_{t_0}^{t}\boldsymbol{H}_o(t,s)\frac{\partial\boldsymbol{u}(s,\tau)}{\partial\tau}\,\mathrm{d}s\\ \displaystyle -\boldsymbol{K}\left\{L_{\boldsymbol{u}}+\boldsymbol{f}_u^{\mathrm{T}}\varphi_{\boldsymbol{x}}+\left(\int_t^{t_f}\boldsymbol{H}_o^{\mathrm{T}}(\sigma,t)\left(L_{\boldsymbol{x}}(\sigma)+\varphi_{tx}(\sigma)+\varphi_{\boldsymbol{xx}}^{\mathrm{T}}(\sigma)\boldsymbol{f}(\sigma)+\boldsymbol{f}_{\boldsymbol{x}}^{\mathrm{T}}(\sigma)\varphi_{\boldsymbol{x}}(\sigma)\right)\mathrm{d}\sigma\right)\right\}\end{bmatrix}, \tag{33}$$

$$\frac{\mathrm{d}t_f}{\mathrm{d}\tau} = -k_{t_f} \left( L + \varphi_t + \varphi_{\boldsymbol{x}}^{\mathrm{T}} \boldsymbol{f} \right)\Big|_{t_f} . \tag{34}$$

Put into this perspective; the initial conditions are

$$\begin{bmatrix} \boldsymbol{x}(t,\tau) \\ \boldsymbol{u}(t,\tau) \end{bmatrix}\Bigg|_{\tau=0} = \begin{bmatrix} \tilde{\boldsymbol{x}}(t) \\ \tilde{\boldsymbol{u}}(t) \end{bmatrix}, \tag{35}$$

$$t_f \big|_{\tau=0} = \tilde{t}_f , \tag{36}$$

where $\tilde{\boldsymbol{x}}(t)$ and $\tilde{\boldsymbol{u}}(t)$ are arbitrary feasible solutions that satisfy Eqs. (11) and (12), and $\tilde{t}_f$ is the initial guess of $t_f$ .

### 3.3 Establishment of Costate-Free Optimality Conditions

The equilibrium solution of the infinite-dimensional dynamic system given by Eqs. (33) and (34) will satisfy

$$L_{\boldsymbol{u}} + \boldsymbol{f}_u^{\mathrm{T}} \varphi_{\boldsymbol{x}} + \left( \int_t^{t_f} \boldsymbol{H}_o^{\mathrm{T}}(\sigma,t) \Big( L_{\boldsymbol{x}}(\sigma) + \varphi_{t\boldsymbol{x}}(\sigma) + \varphi_{\boldsymbol{xx}}^{\mathrm{T}}(\sigma) \boldsymbol{f}(\sigma) + \boldsymbol{f}_{\boldsymbol{x}}^{\mathrm{T}}(\sigma) \varphi_{\boldsymbol{x}}(\sigma) \Big) \mathrm{d}\sigma \right) = \boldsymbol{0} , \tag{37}$$

$$L(t_f) + \phi_t(t_f) + \varphi_{\boldsymbol{x}}^{\mathrm{T}}(t_f) \boldsymbol{f}(t_f) = 0 . \tag{38}$$

Actually, Eqs. (37) and (38) are the first-order optimality conditions for Problem 2 without the employment of costates. We will show that they are equivalent to the classic ones with costates [33], i.e.,

$$\dot{\boldsymbol{\lambda}} + H_{\boldsymbol{x}} = \dot{\boldsymbol{\lambda}} + L_{\boldsymbol{x}} + \boldsymbol{f}_{\boldsymbol{x}}^{\mathrm{T}} \boldsymbol{\lambda} = \boldsymbol{0} , \tag{39}$$

$$H_{\boldsymbol{u}} = L_{\boldsymbol{u}} + \boldsymbol{f}_{\boldsymbol{u}}^{\mathrm{T}} \boldsymbol{\lambda} = \boldsymbol{0} , \tag{40}$$

and the transversality conditions

$$\boldsymbol{\lambda}(t_f) = \varphi_{\boldsymbol{x}}(t_f) , \tag{41}$$

$$H(t_f) + \varphi_{t_f} = 0 , \tag{42}$$

where $H = L + \boldsymbol{\lambda}^{\mathrm{T}} \boldsymbol{f}$ is the Hamiltonian and $\boldsymbol{\lambda}$ is the costate vector.

**Theorem 3.1** *For Problem 2, the optimality conditions given by Eqs.* (37) *and* (38) *are equivalent to the optimality conditions given by Eqs.* (39)-(42).

***Proof*** With Eq. (19), Eq. (37) may be reformulated as

$$L_{\boldsymbol{u}} + \boldsymbol{f}_u^{\mathrm{T}} \varphi_{\boldsymbol{x}} + \boldsymbol{f}_u^{\mathrm{T}} \left( \int_t^{t_f} \boldsymbol{\Phi}_o^{\mathrm{T}}(\sigma,t) \Big( L_{\boldsymbol{x}}(\sigma) + \varphi_{t\boldsymbol{x}}(\sigma) + \varphi_{\boldsymbol{xx}}^{\mathrm{T}}(\sigma) \boldsymbol{f}(\sigma) + \boldsymbol{f}_{\boldsymbol{x}}^{\mathrm{T}}(\sigma) \varphi_{\boldsymbol{x}}(\sigma) \Big) \mathrm{d}\sigma \right) = \boldsymbol{0} , \tag{43}$$

where $\boldsymbol{\Phi}_o(\sigma,t)$ is the state transition matrix related to $\boldsymbol{H}_o(\sigma,t)$ . Define a quantity $\boldsymbol{\gamma}(t)$ as

$$\boldsymbol{\gamma}(t) = \varphi_{\boldsymbol{x}}(t) + \int_t^{t_f} \boldsymbol{\Phi}_o^{\mathrm{T}}(\sigma,t) \Big( L_{\boldsymbol{x}}(\sigma) + \varphi_{t\boldsymbol{x}}(\sigma) + \varphi_{\boldsymbol{xx}}^{\mathrm{T}}(\sigma) \boldsymbol{f}(\sigma) + \boldsymbol{f}_{\boldsymbol{x}}^{\mathrm{T}}(\sigma) \varphi_{\boldsymbol{x}}(\sigma) \Big) \mathrm{d}\sigma . \tag{44}$$

Then Eq. (43) is simplified as

$$L_{\boldsymbol{u}}+\boldsymbol{f_u}^{\mathrm{T}}\boldsymbol{\gamma}=\boldsymbol{0}\ . \tag{45}$$

Obviously, from Eq. (44), when $t=t_f$, there is

$$\boldsymbol{\gamma}(t_f)=\phi_{\boldsymbol{x}}(t_f)\ . \tag{46}$$

Differentiate $\boldsymbol{\gamma}(t)$ with respect to $t$. In the process, we will use the Leibniz rule [34]

$$\frac{\mathrm{d}}{\mathrm{d}t}\left(\int_{b(t)}^{a(t)}h(\sigma,t)\,\mathrm{d}\sigma\right)=h\big(a(t),t\big)\frac{\mathrm{d}}{\mathrm{d}t}a(t)-h\big(b(t),t\big)\frac{\mathrm{d}}{\mathrm{d}t}b(t)+\int_{b(t)}^{a(t)}h_t(\sigma,t)\,\mathrm{d}\sigma\ , \tag{47}$$

and the property of $\boldsymbol{\Phi}_o(\sigma,t)$ (see Lemma 3.1)

$$\frac{\partial\boldsymbol{\Phi}_o(\sigma,t)}{\partial t}=-\boldsymbol{\Phi}_o(\sigma,t)\boldsymbol{f_x}(t)\ , \tag{48}$$

$$\boldsymbol{\Phi}_o(t,t)=\boldsymbol{I}_{n\times n}\ . \tag{49}$$

Then we have

$$\begin{aligned}\frac{\mathrm{d}}{\mathrm{d}t}\boldsymbol{\gamma}(t)&=\varphi_{t\boldsymbol{x}}+\varphi_{\boldsymbol{xx}}{}^{\mathrm{T}}\boldsymbol{f}-\left(L_{\boldsymbol{x}}+\varphi_{t\boldsymbol{x}}+\varphi_{\boldsymbol{xx}}{}^{\mathrm{T}}\boldsymbol{f}+\boldsymbol{f_x}^{\mathrm{T}}\varphi_{\boldsymbol{x}}\right)-\boldsymbol{f_x}^{\mathrm{T}}\int_t^{t_f}\boldsymbol{\Phi}_o{}^{\mathrm{T}}(\sigma,t)\Big(L_{\boldsymbol{x}}(\sigma)+\varphi_{t\boldsymbol{x}}(\sigma)+\varphi_{\boldsymbol{xx}}{}^{\mathrm{T}}(\sigma)\boldsymbol{f}(\sigma)+\boldsymbol{f_x}^{\mathrm{T}}(\sigma)\varphi_{\boldsymbol{x}}(\sigma)\Big)\mathrm{d}\sigma\\&=-L_{\boldsymbol{x}}-\boldsymbol{f_x}^{\mathrm{T}}\left(\varphi_{\boldsymbol{x}}(t)+\int_t^{t_f}\boldsymbol{\Phi}_o{}^{\mathrm{T}}(\sigma,t)\Big(L_{\boldsymbol{x}}(\sigma)+\varphi_{t\boldsymbol{x}}(\sigma)+\varphi_{\boldsymbol{xx}}{}^{\mathrm{T}}(\sigma)\boldsymbol{f}(\sigma)+\boldsymbol{f_x}^{\mathrm{T}}(\sigma)\varphi_{\boldsymbol{x}}(\sigma)\Big)\mathrm{d}\sigma\right)\\&=-L_{\boldsymbol{x}}-\boldsymbol{f_x}^{\mathrm{T}}\boldsymbol{\gamma}(t)\end{aligned} \tag{50}$$

From Eqs. (50) and (46), it is found that $\boldsymbol{\gamma}(t)$ conforms to the same dynamics and boundary conditions as the costates $\boldsymbol{\lambda}(t)$. Thus, we can conclude that $\boldsymbol{\gamma}(t)=\boldsymbol{\lambda}(t)$. Then Eq. (45) and Eq. (40) are identical. This means that Eq. (37) is equivalent to Eqs. (39), (40) and (41). On the other hand, with Eq. (41), it is easy to show that Eqs. (38) and (42) are the same. □

According to the above analysis, the costate-free Hamiltonian may be presented as

$$H=L+\boldsymbol{f}^{\mathrm{T}}\left(\varphi_{\boldsymbol{x}}(t)+\int_t^{t_f}\boldsymbol{\Phi}_o{}^{\mathrm{T}}(\sigma,t)\Big(L_{\boldsymbol{x}}(\sigma)+\varphi_{t\boldsymbol{x}}(\sigma)+\varphi_{\boldsymbol{xx}}{}^{\mathrm{T}}(\sigma)\boldsymbol{f}(\sigma)+\boldsymbol{f_x}^{\mathrm{T}}(\sigma)\varphi_{\boldsymbol{x}}(\sigma)\Big)\mathrm{d}\sigma\right). \tag{51}$$

Then the EPDE for the control variables is simply reformulated as

$$\frac{\partial\boldsymbol{u}(t,\tau)}{\partial\tau}=-\boldsymbol{K}H_{\boldsymbol{u}}\ . \tag{52}$$

An interesting trial on the costate-free optimality conditions is to solve the optimal control from these theoretical expressions. By investigating Eq. (37), it is found that the optimal control is related to the future states. This means generally the optimal feedback control law in the analytic form cannot be solved. However, for the linear quadratic regulator (LQR) problem with the state equation given by Eq. (17) and the performance index given by

$$J = \frac{1}{2}\int_{t_0}^{\infty} (\boldsymbol{x}^{\mathrm{T}}\boldsymbol{Q}\boldsymbol{x} + \boldsymbol{u}^{\mathrm{T}}\boldsymbol{R}\boldsymbol{u})\,\mathrm{d}t\,, \tag{53}$$

where $\boldsymbol{Q}$ and $\boldsymbol{R}$ are positive-definite matrixes, the optimal control law may be derived as

$$\boldsymbol{u}(t) = -\boldsymbol{R}^{-1}\boldsymbol{B}^{\mathrm{T}}(t)\int_{t}^{\infty} \boldsymbol{\Phi}^{\mathrm{T}}(\sigma,t)\boldsymbol{Q}\boldsymbol{x}(\sigma)\,\mathrm{d}\sigma\,. \tag{54}$$

subject to

$$\dot{\boldsymbol{x}} = \boldsymbol{A}(t)\boldsymbol{x} - \boldsymbol{B}(t)\boldsymbol{R}^{-1}\boldsymbol{B}^{\mathrm{T}}(t)\int_{t}^{\infty} \boldsymbol{\Phi}^{\mathrm{T}}(\sigma,t)\boldsymbol{Q}\boldsymbol{x}(\sigma)\,\mathrm{d}\sigma\,, \tag{55}$$

and the closed-form analytic solution may be obtained upon the following identity

$$\int_{t}^{\infty} \boldsymbol{\Phi}^{\mathrm{T}}(\sigma,t)\boldsymbol{Q}\boldsymbol{x}(\sigma)\,\mathrm{d}\sigma = \boldsymbol{P}(t)\boldsymbol{x}(t)\,, \tag{56}$$

where the matrix $\boldsymbol{P}(t)$ is the solution of the Riccati differential equation

$$\dot{\boldsymbol{P}} = -\boldsymbol{A}^{\mathrm{T}}\boldsymbol{P} - \boldsymbol{P}\boldsymbol{A} + \boldsymbol{P}\boldsymbol{B}\boldsymbol{R}^{-1}\boldsymbol{B}^{\mathrm{T}}\boldsymbol{P} - \boldsymbol{Q}\,,\ \boldsymbol{P}(+\infty) = \boldsymbol{0}_{n\times n}\,. \tag{57}$$

This may be verified by eliminating $\boldsymbol{x}(t)$ in Eq. (56) and then implementing the differentiation. When the state equation (17) is time-invariant, then the closed-form analytic solution may be obtained upon the following identity

$$\int_{t}^{\infty} e^{\boldsymbol{A}^{\mathrm{T}}(\sigma-t)}\boldsymbol{Q}\boldsymbol{x}(\sigma)\,\mathrm{d}\sigma = \boldsymbol{P}\boldsymbol{x}(t)\,, \tag{58}$$

where the matrix $\boldsymbol{P}$ is the solution of the Riccati algebraic equation

$$\boldsymbol{A}^{\mathrm{T}}\boldsymbol{P} + \boldsymbol{P}\boldsymbol{A} - \boldsymbol{P}\boldsymbol{B}\boldsymbol{R}^{-1}\boldsymbol{B}^{\mathrm{T}}\boldsymbol{P} + \boldsymbol{Q} = \boldsymbol{0}_{n\times n}\,. \tag{59}$$

After proving the equivalence of optimality conditions, now the variables' evolving direction using the VEM is easy to determine.

**Theorem 3.2** *Solving the IVP defined by Eqs.* (33)-(36)*, when* $\tau \to +\infty$*,* $(\boldsymbol{x},\boldsymbol{u},t_f)$ *will satisfy the optimality conditions of Problem 2.*

***Proof*** Upon the infinite-dimensional dynamics governed by Eqs. (33) and (34), with the initial conditions of Eqs. (35) and (36), we have $\frac{\delta J}{\delta \tau} \le 0$ and $\frac{\delta J}{\delta \tau} = 0$ when $J$ reaches its minimum within the feasible solution domain $\mathbb{D}_o$. That is, the functional (10) is the Lyapunov functional in $\mathbb{D}_o$. According to Lemma 2.1, the minimum solution of Eq. (10) is an asymptotically stable equilibrium solution. Thus, the solution of the IVP will asymptotically meet the optimality conditions, namely, Eqs. (37) and (38). □

Note that although only the first-order optimality conditions are explicitly guaranteed, the solution of the IVP, defined by Eqs. (33)-(36), will not halt at a maximum or a saddle point of Problem 2 (unless they are the initial guesses), since those solutions are not asymptotically stable equilibrium solutions.

Presume that we already have a feasible initial solution of $\tilde{\boldsymbol{x}}(t)$ and $\tilde{\boldsymbol{u}}(t)$; then theoretically Eqs. (33) and (34) may be used to obtain the optimal solution of Problem 2. Recall Fig. 1, Eqs. (33) and (34) realize the anticipated variable evolution along the variation time $\tau$. The initial conditions of $\boldsymbol{x}(t,\tau)$ and $\boldsymbol{u}(t,\tau)$ at $\tau=0$ belong to the feasible solution domain and their value at $\tau=+\infty$ is optimal for the OCP. The right part of the EPDE (33) is also only a vector function of time $t$. Thus we may apply the semi-discrete method to discretize it along the normal time dimension and further use ODE integration methods to get the numerical solution. The OCP defined in Problem 2 has a free terminal time $t_f$. When $t_f$ is fixed, the evolution equations for the variables $\boldsymbol{x}$ and $\boldsymbol{u}$ remain the same, while the equation regarding the terminal time $t_f$ is no longer necessary.

### 3.4 Computation of Impulse Response Function

In employing the EPDE (33) to get the solution, one important issue is to determine the impulse response function $\boldsymbol{H}_o(t,s)$, which is related to the current solution of $\boldsymbol{x}(t)$ and $\boldsymbol{u}(t)$. We have

$$\boldsymbol{H}_o(t,s)=\boldsymbol{0}_{n\times m} \quad \textit{when} \;\; t<s\,, \tag{60}$$

and for $t\geq s$, there is

$$\frac{\mathrm{d}}{\mathrm{d}t}\boldsymbol{H}_o(t,s)=\boldsymbol{f}_{\boldsymbol{x}}(t)\boldsymbol{H}_o(t,s),\;\; \boldsymbol{H}_o(t,s)=\boldsymbol{f}_{\boldsymbol{u}}(s) \;\; \textit{when} \;\; t=s\,. \tag{61}$$

When applying the semi-discrete method for the initial-value problem of the PDE system, the numerical computation of $\boldsymbol{H}_o(t,s)$, along the discretization of the time coordinate, i.e., $t_i$ $(i=1,2,...,N)$ with $N$ being the number of discretization time points, is performed as follows.

Given a fixed time point $t_j$, for any $t_i>t_j$, we may use an implicit integration format, which brings higher precision and a larger stability region than those of the explicit format, to get

$$\begin{bmatrix}\boldsymbol{H}_o(t_{j+1},t_j)\\ \boldsymbol{H}_o(t_{j+2},t_j)\\ \cdots\\ \boldsymbol{H}_o(t_N,t_j)\end{bmatrix}=\begin{bmatrix}\boldsymbol{H}_o(t_j,t_j)\\ \boldsymbol{H}_o(t_{j+1},t_j)\\ \cdots\\ \boldsymbol{H}_o(t_{N-1},t_j)\end{bmatrix}+\begin{bmatrix}0.5(t_{j+1}-t_j)\left(\boldsymbol{f}_{\boldsymbol{x}}(t_{j+1})\boldsymbol{H}_o(t_{j+1},t_j)+\boldsymbol{f}_{\boldsymbol{x}}(t_j)\boldsymbol{H}_o(t_j,t_j)\right)\\ 0.5(t_{j+2}-t_{j+1})\left(\boldsymbol{f}_{\boldsymbol{x}}(t_{j+2})\boldsymbol{H}_o(t_{j+2},t_j)+\boldsymbol{f}_{\boldsymbol{x}}(t_{j+1})\boldsymbol{H}_o(t_{j+1},t_j)\right)\\ \cdots\\ 0.5(t_N-t_{N-1})\left(\boldsymbol{f}_{\boldsymbol{x}}(t_N)\boldsymbol{H}_o(t_N,t_j)+\boldsymbol{f}_{\boldsymbol{x}}(t_{N-1})\boldsymbol{H}_o(t_{N-1},t_j)\right)\end{bmatrix}. \tag{62}$$

Then we may derive that

$$\boldsymbol{Mtx}_{\boldsymbol{H}_o}^{j+1} = \left(\boldsymbol{C}_1^{\ j+1} - \boldsymbol{C}_2^{\ j+1}\boldsymbol{C}_3^{\ j+1}\boldsymbol{C}_4^{\ j+1}\right)^{-1} \begin{bmatrix} \left(0.5(t_{j+1}-t_j)\boldsymbol{f}_{\boldsymbol{x}}(t_j) + \boldsymbol{I}_{n\times n}\right) \\ \boldsymbol{0}_{n\times n} \\ \cdots \\ \boldsymbol{0}_{n\times n} \end{bmatrix} \boldsymbol{H}_o(t_j, t_j) \,. \tag{63}$$

where

$$\boldsymbol{Mtx}_{\boldsymbol{H}_o}^{j+1} = \begin{bmatrix} \boldsymbol{H}_o(t_{j+1}, t_j) \\ \boldsymbol{H}_o(t_{j+2}, t_j) \\ \cdots \\ \boldsymbol{H}_o(t_N, t_j) \end{bmatrix}, \quad \boldsymbol{C}_1^{\ j} = \begin{bmatrix} \boldsymbol{I}_{n\times n} & & & \\ -\boldsymbol{I}_{n\times n} & \boldsymbol{I}_{n\times n} & & \\ & \cdots & \cdots & \\ & & -\boldsymbol{I}_{n\times n} & \boldsymbol{I}_{n\times n} \end{bmatrix}, \quad \boldsymbol{C}_3^{\ j} = \begin{bmatrix} \boldsymbol{I}_{n\times n} & & & \\ \boldsymbol{I}_{n\times n} & \boldsymbol{I}_{n\times n} & & \\ & \cdots & \boldsymbol{I}_{n\times n} & \\ & & \boldsymbol{I}_{n\times n} & \boldsymbol{I}_{n\times n} \end{bmatrix},$$

$$\boldsymbol{C}_2^{\ j} = \begin{bmatrix} 0.5(t_{j+1}-t_j)\boldsymbol{I}_{n\times n} & & & \\ & 0.5(t_{j+2}-t_{j+1})\boldsymbol{I}_{n\times n} & & \\ & & \cdots & \\ & & & 0.5(t_N - t_{N-1})\boldsymbol{I}_{n\times n} \end{bmatrix}, \text{ and } \boldsymbol{C}_4^{\ j} = \begin{bmatrix} \boldsymbol{f}_{\boldsymbol{x}}(t_{j+1}) & & & \\ & \boldsymbol{f}_{\boldsymbol{x}}(t_{j+2}) & & \\ & & \cdots & \\ & & & \boldsymbol{f}_{\boldsymbol{x}}(t_N) \end{bmatrix}.$$

Note that $(\boldsymbol{C}_1^{\ j+1} - \boldsymbol{C}_2^{\ j+1}\boldsymbol{C}_3^{\ j+1}\boldsymbol{C}_4^{\ j+1})$ is a triangular matrix and its inverse always exists for a sufficiently small discretization granularity. Therefore, the values of the impulse response function at different time points, with respect to $t_j$, are

$$\begin{bmatrix} \boldsymbol{H}_o(t_1, t_j) \\ \cdots \\ \boldsymbol{H}_o(t_j, t_j) \\ \boldsymbol{H}_o(t_{j+1}, t_j) \\ \cdots \\ \boldsymbol{H}_o(t_N, t_j) \end{bmatrix} = \begin{bmatrix} \boldsymbol{0}_{n\times n} \\ \cdots \\ \boldsymbol{I}_{n\times n} \\ \left(\boldsymbol{C}_1^{\ j+1} - \boldsymbol{C}_2^{\ j+1}\boldsymbol{C}_3^{\ j+1}\boldsymbol{C}_4^{\ j+1}\right)^{-1} \begin{bmatrix} \left(0.5(t_{j+1}-t_j)\boldsymbol{f}_{\boldsymbol{x}}(t_j) + \boldsymbol{I}_n\right) \\ \boldsymbol{0}_{n\times n} \\ \cdots \\ \boldsymbol{0}_{n\times n} \end{bmatrix} \end{bmatrix} \boldsymbol{f}_{\boldsymbol{u}}(t_j) \,. \tag{64}$$

# 4 Examples

First, a linear example taken from Xie [35] is considered.

*Example 4.1* Consider the following dynamic system

$$\dot{\boldsymbol{x}} = \boldsymbol{A}\boldsymbol{x} + \boldsymbol{b}u \,,$$

where $\boldsymbol{x} = \begin{bmatrix} x_1 \\ x_2 \end{bmatrix}$, $\boldsymbol{A} = \begin{bmatrix} 0 & 1 \\ 0 & 0 \end{bmatrix}$, and $\boldsymbol{b} = \begin{bmatrix} 0 \\ 1 \end{bmatrix}$. Find the solution that minimizes the performance index

$$J = \frac{1}{2}\boldsymbol{x}(t_f)^{\mathrm{T}}\boldsymbol{F}\boldsymbol{x}(t_f) + \frac{1}{2}\int_{t_0}^{t_f}\left(\boldsymbol{x}^{\mathrm{T}}\boldsymbol{Q}\boldsymbol{x} + Ru^2\right)\mathrm{d}t \,,$$

with the initial boundary conditions $\boldsymbol{x}(t_0)=\begin{bmatrix}1\\1\end{bmatrix}$, where the initial time $t_0=0$ and the terminal time $t_f=3$ are fixed. The weighted matrixes are $\boldsymbol{F}=\begin{bmatrix}1 & 0\\0 & 2\end{bmatrix}$, $\boldsymbol{Q}=\begin{bmatrix}2 & 1\\1 & 4\end{bmatrix}$ and $R=\dfrac{1}{2}$.

In solving this example using the VEM, the EPDE derived was

$$\frac{\partial}{\partial\tau}\begin{bmatrix}\boldsymbol{x}\\u\end{bmatrix}=\begin{bmatrix}\int_{t_0}^{t}e^{\boldsymbol{A}(t-s)}\boldsymbol{b}\dfrac{\partial u}{\partial\tau}(s)\,\mathrm{d}s\\-K\left\{Ru+\boldsymbol{b}^{\mathrm{T}}\boldsymbol{F}\boldsymbol{x}+\boldsymbol{b}^{\mathrm{T}}\left(\int_{t}^{t_f}e^{\boldsymbol{A}^{\mathrm{T}}(\sigma-t)}\left((\boldsymbol{Q}+\boldsymbol{F}\boldsymbol{A}+\boldsymbol{A}^{\mathrm{T}}\boldsymbol{F})\boldsymbol{x}(\sigma)+\boldsymbol{F}\boldsymbol{b}u(\sigma)\right)\mathrm{d}\sigma\right)\right\}\end{bmatrix},$$

with the one-dimensional gain $K$ set as $K=2\times10^{-2}$. The initial conditions of the EPDE, with the control input $\tilde{u}(t)=0$, were $\tilde{\boldsymbol{x}}(t)=\begin{bmatrix}t+1\\1\end{bmatrix}$. Using the semi-discrete method, the time horizon $[t_0,\, t_f]$ was discretized uniformly with 61 points. Thus, a dynamic system with 183 states was obtained and the OCP was transformed to a finite-dimensional IVP. The ODE integrator "ode45" in MATLAB, with a default relative error tolerance of $1\times10^{-3}$ and a default absolute error tolerance of $1\times10^{-6}$, was employed to solve the IVP. Even for this simple example, the analytic solution is not easy to obtain. For comparison, we computed the optimal solution with GPOPS-II [36], a Radau PS method-based OCP solver.

Figures 2, 3 and 4 show the evolution process of $x_1(t)$, $x_2(t)$ and $u(t)$ solutions to their optimal values, respectively. At $\tau$ = 300 s, the numerical solutions are indistinguishable from the optimal, which shows the effectiveness of the VEM. Figure 5 plots the profile of the performance index value against the variation time. This index declines rapidly at first and is close to the minimum when $\tau$ = 50 s. Then it approaches the minimum monotonically and is almost fixed at the optimal value.

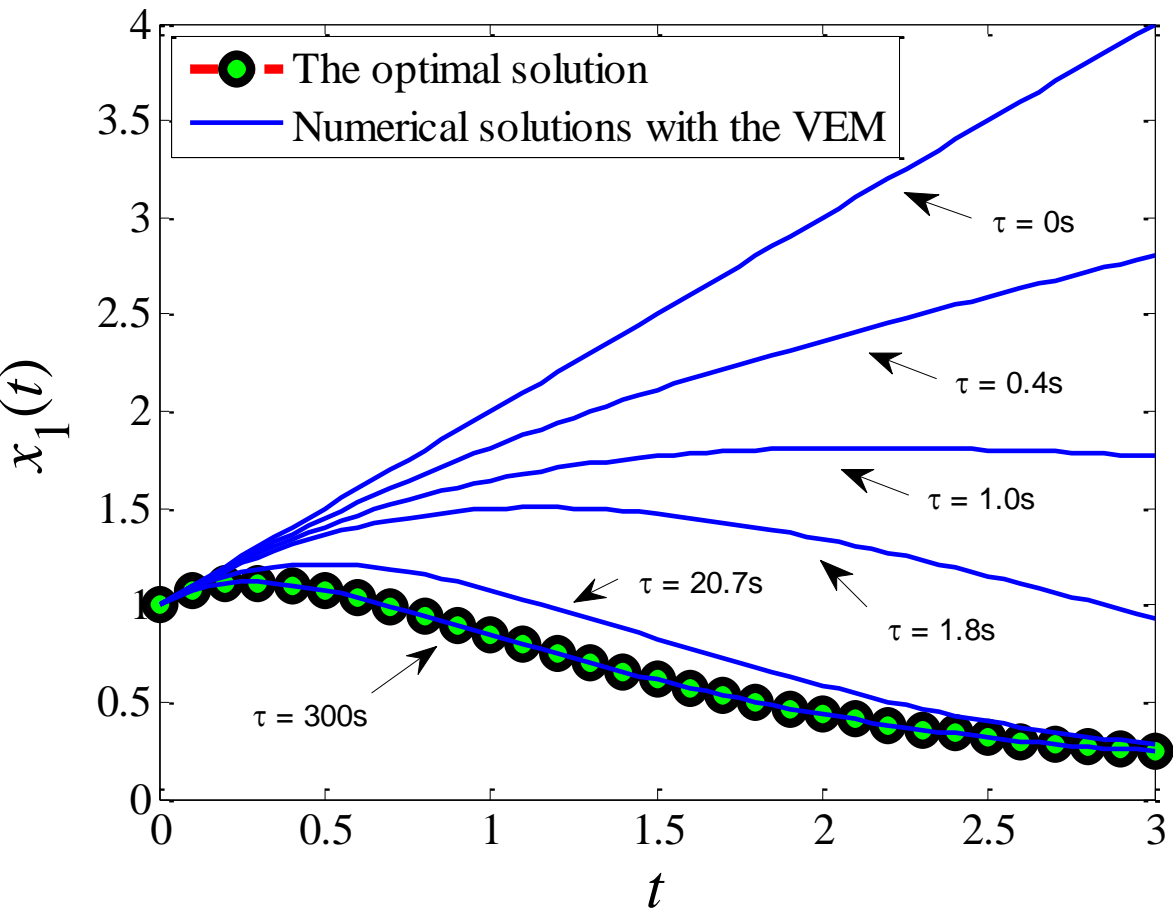

**Fig. 2** The evolution of numerical solutions of $x_1$ to the optimal solution

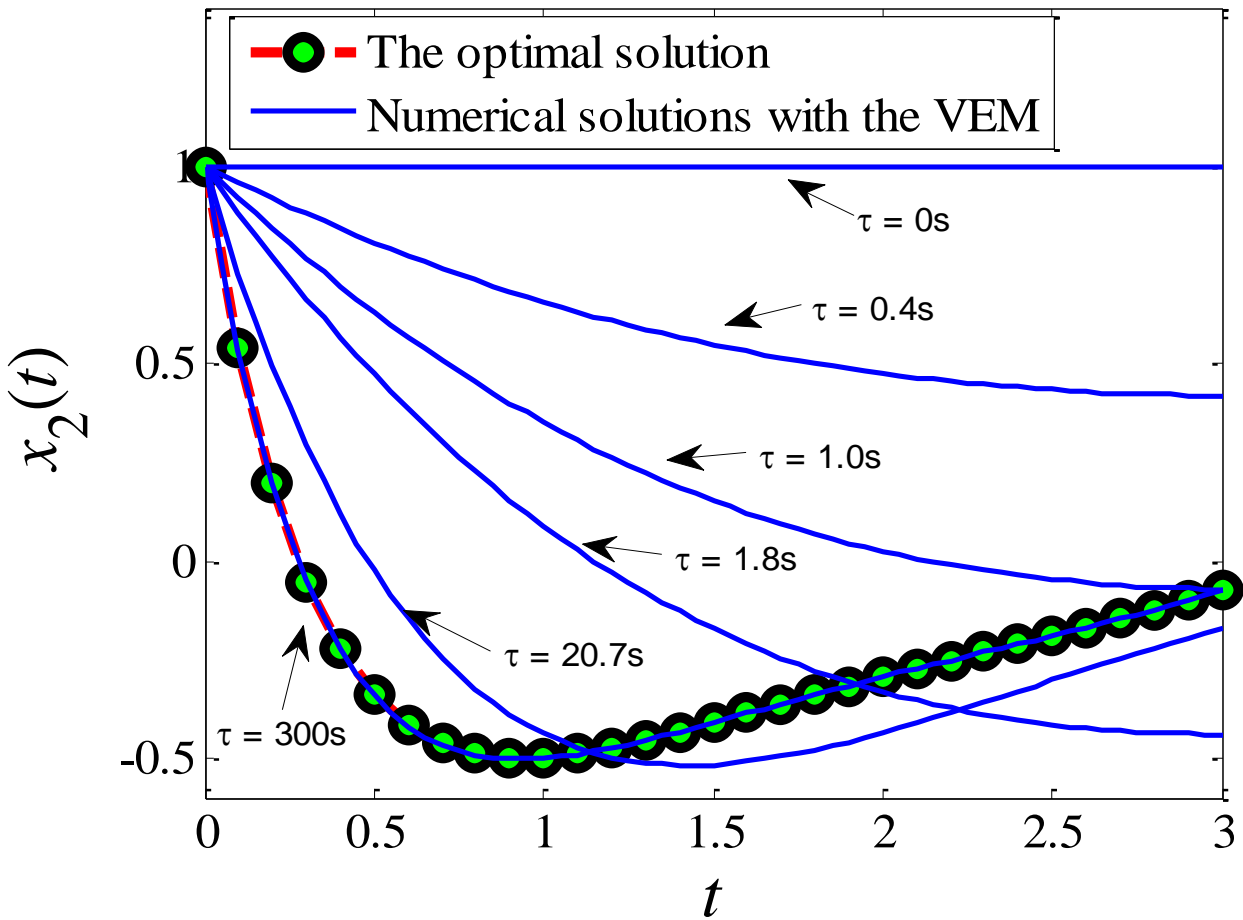

**Fig. 3** The evolution of numerical solutions of $x_2$ to the optimal solution

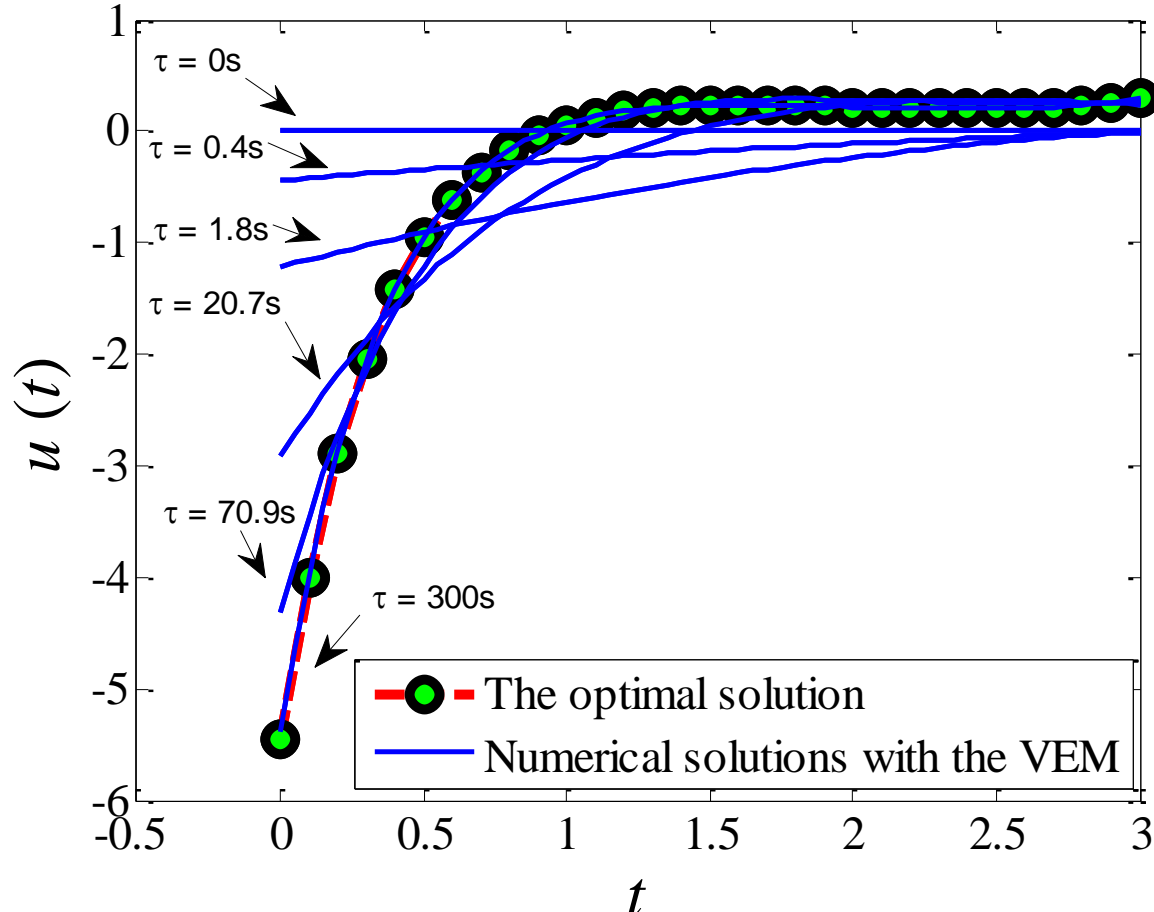

**Fig. 4** The evolution of numerical solutions of $u$ to the optimal solution

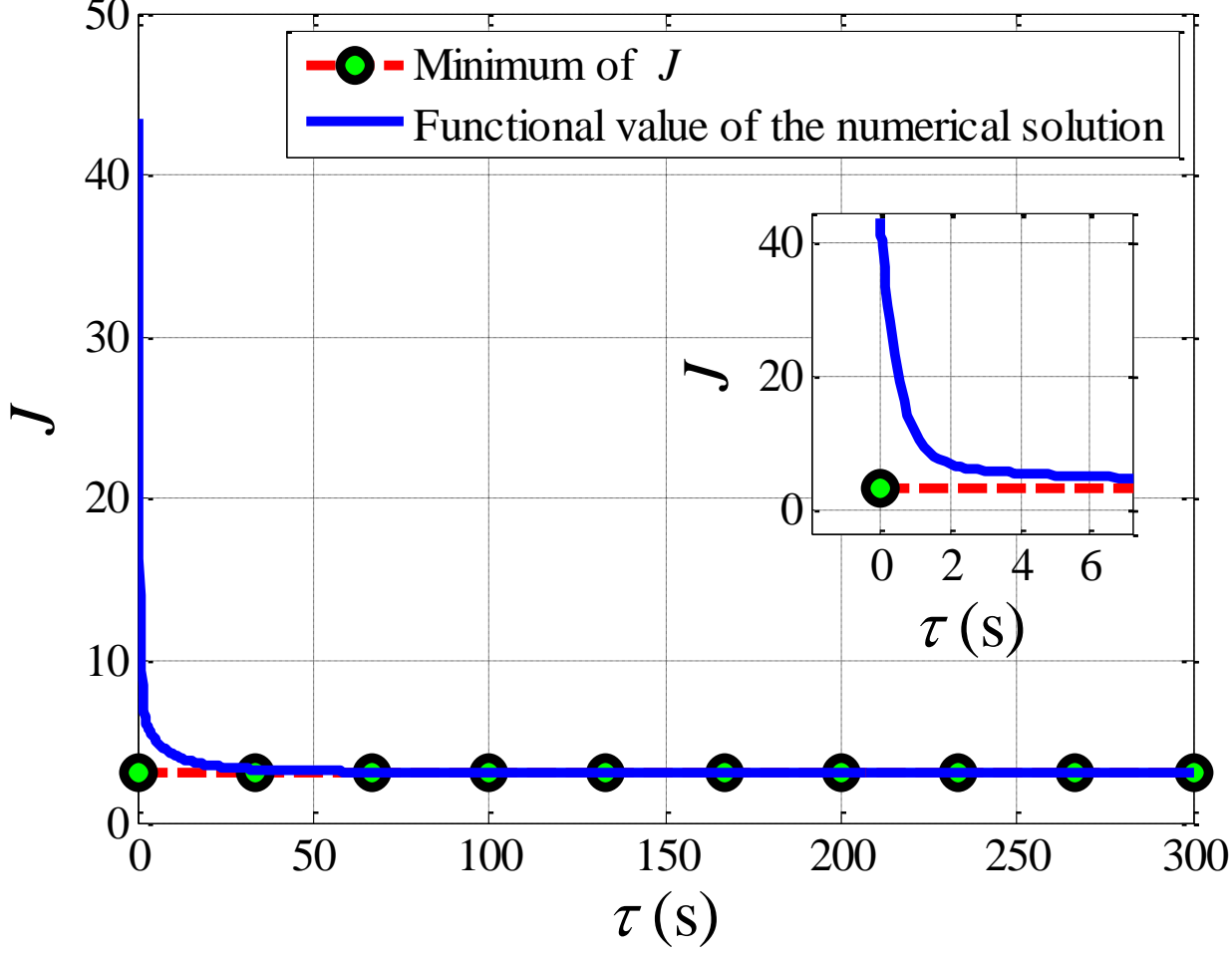

**Fig. 5** The approach to the minimum performance index

Now we consider a nonlinear example with free terminal time $t_f$, the 2-dimensional homing missile problem adapted from Hull [37].

*Example 4.2* Consider the problem of a constant speed missile intercepting a constant speed target moving in a straight line. The dynamic equations are

$$\dot{\boldsymbol{x}} = \boldsymbol{f}(\boldsymbol{x}, u),$$

where $\boldsymbol{x} = \begin{bmatrix} x \\ y \\ \theta_M \end{bmatrix}$ and $\boldsymbol{f} = \begin{bmatrix} V_T \cos(\theta_T) - V_M \cos(\theta_M) \\ V_T \sin(\theta_T) - V_M \sin(\theta_M) \\ \dfrac{u}{V_M} \end{bmatrix}$. Here, $x$ and $y$ are the abscissa and ordinate of the relative position, respectively, $\theta_M$ is the azimuth of the missile, $\theta_T$ =30 deg is the azimuth of the target, $V_M$ =1000 m/s is the constant speed of the missile, $V_T$ =500 m/s is the constant speed of the target, and $u$ is the missile normal acceleration. To intercept the target and penalize too large normal acceleration, the performance index to be minimized is defined as

$$J = \frac{1}{2}\boldsymbol{x}(t_f)^{\mathrm{T}} \boldsymbol{F} \boldsymbol{x}(t_f) + \frac{1}{2}\int_{t_0}^{t_f} R u^2 \mathrm{d}t,$$

where the weighted matrixes are $\boldsymbol{F} = \begin{bmatrix} 1\times10^{-2} & 0 & 0 \\ 0 & 2\times10^{-2} & 0 \\ 0 & 0 & 0 \end{bmatrix}$ and $R = 5\times10^{-4}$. The initial boundary conditions are $\left.\begin{bmatrix} x \\ y \\ \theta_M \end{bmatrix}\right|_{t_0=0} = \begin{bmatrix} 10000\mathrm{m} \\ 5000\mathrm{m} \\ 0\,\mathrm{deg} \end{bmatrix}$ and the terminal time $t_f$ is free.

Before the computation, the state and the control variables were scaled to improve the numerical efficiency. In the specific form of the EPDE (33) and the EDE(34), the gain parameters $K$ and $k_{t_f}$ were set to be $1.5\times10^{-6}$ and $1\times10^{-4}$, respectively. The initial conditions for the evolution equations, i.e., $\left.\begin{bmatrix} \boldsymbol{x}(t,\tau) \\ u(t,\tau) \\ t_f(\tau) \end{bmatrix}\right|_{\tau=0}$, were obtained by numerical integration at the time horizon of [0, 25]s with the control input $\tilde{u}(t) = 0$. We also discretized the time horizon $[t_0, t_f]$ uniformly, with 51 points. Therefore, an IVP with 205 states (including the terminal time) was obtained. We still employed "ode45" in MATLAB for the numerical integration. In the integrator setting, the default relative error tolerance and the absolute error tolerance were $1\times10^{-3}$ and $1\times10^{-6}$, respectively. For comparison, the optimal solution was again computed with GPOPS-II.

Figure 6 gives the states curve in the $xy$ relative position coordinate plane, showing that the numerical results tend to the optimal solution over time. For the optimal solution, the missile will intercept the target with a fairly small position error. The control solutions are plotted in Fig. 7, and the asymptotical approach of the numerical results to the optimal solution is demonstrated. In Fig. 8, the terminal time profile against the variation time $\tau$ is plotted. The results of $t_f$ oscillate sharply at first and then gradually approach the optimal interception time, and the value only changes slightly after $\tau = 40$ s. At $\tau = 300$ s, we compute that $t_f = 23.52$ s from the VEM, the same as the result from GPOPS-II.

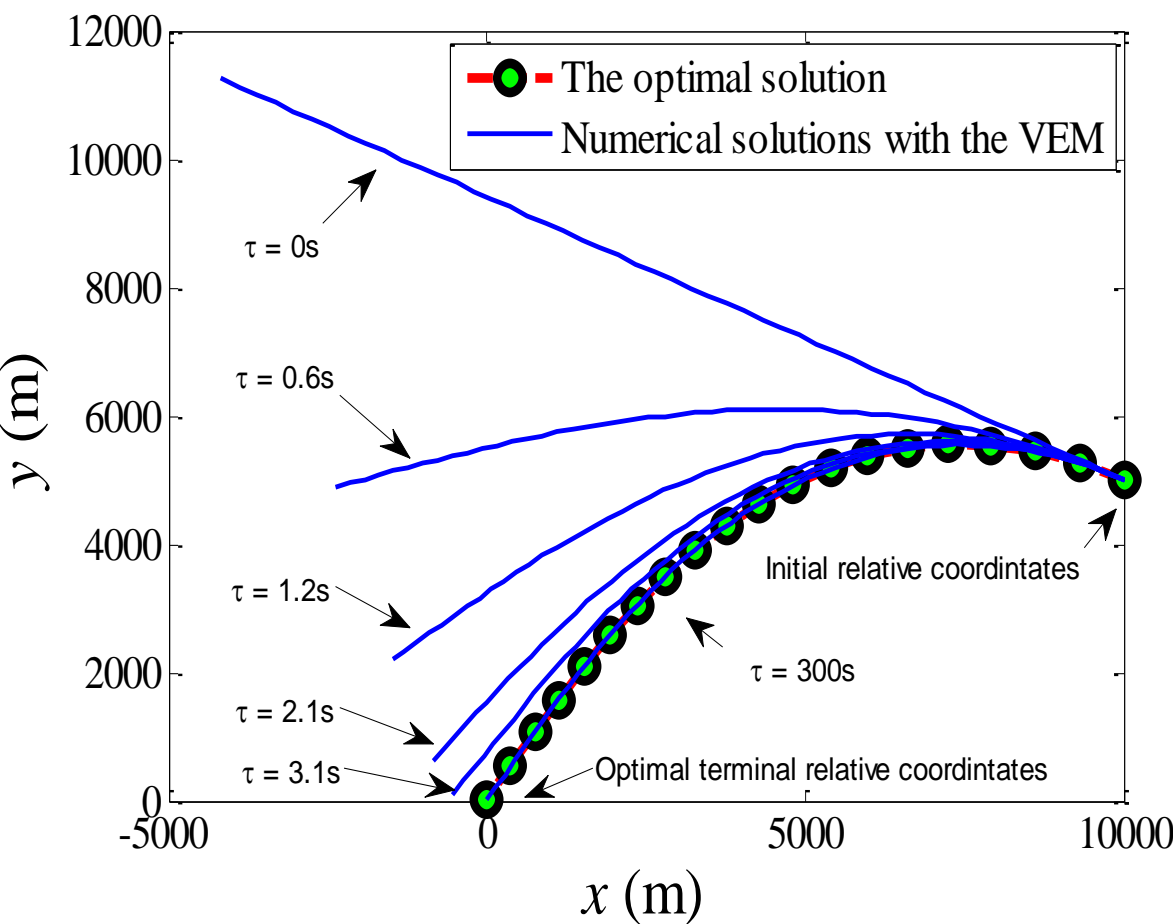

**Fig. 6** The evolution of numerical solutions in the $xy$ coordinate plane to the optimal solution

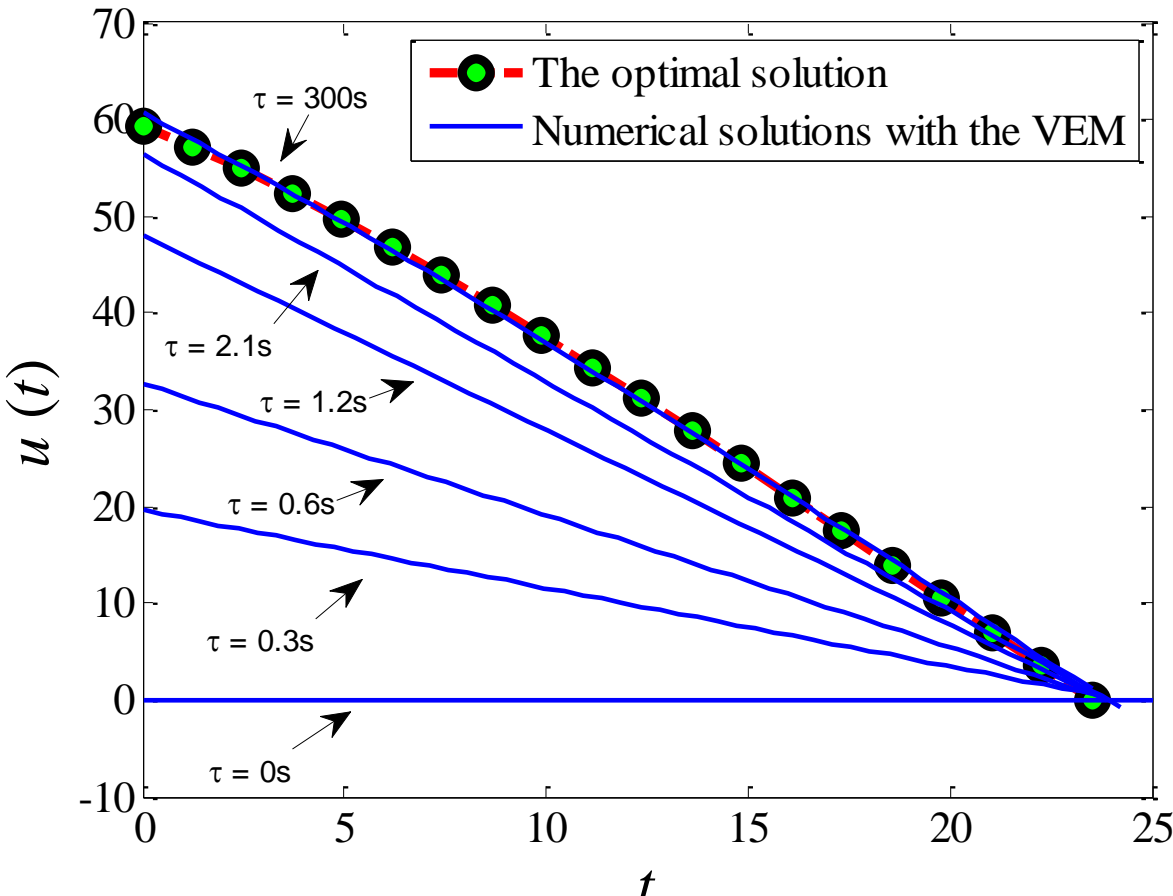

**Fig. 7** The evolution of numerical solutions of $u$ to the optimal solution

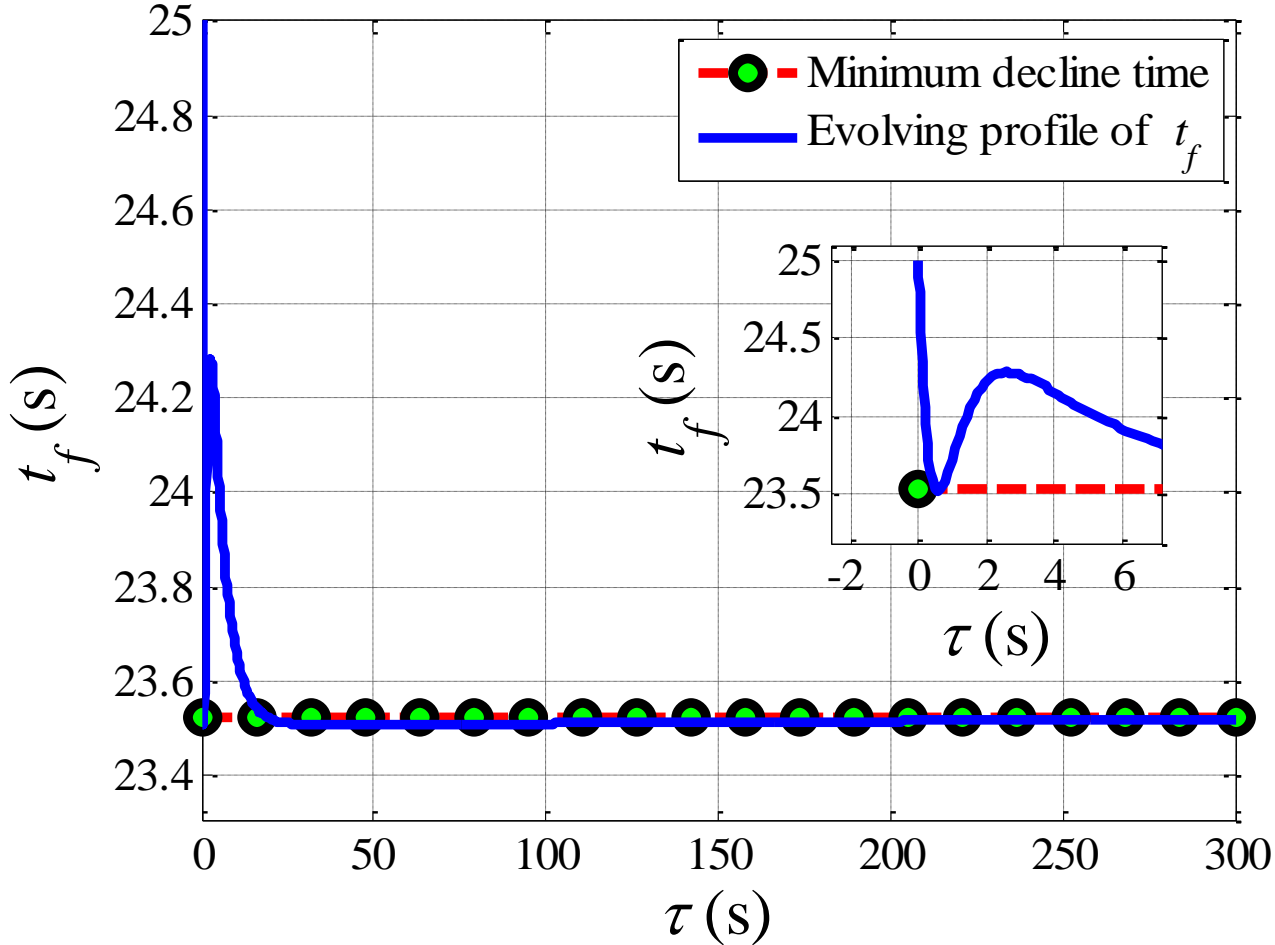

**Fig. 8** The evolution profile of $t_f$ to the optimal interception time

# 5 Further Comments

In contrast to the augmented EPDE (AEPDE) that introduced the costates in Ref. [25], this paper develops a compact version of the EPDE that uses the primal variables only. Discussion between the two formulations is worthwhile. For convenience, the AEPDE is again presented.

$$\frac{\partial \boldsymbol{y}(t,\tau)}{\partial \tau} = -\boldsymbol{K}\left( H_{yy} \begin{bmatrix} \left( H_x + \dfrac{\partial \boldsymbol{\lambda}}{\partial t} \right) \\ \left( \boldsymbol{f} - \dfrac{\partial \boldsymbol{x}}{\partial t} \right) \\ H_u \end{bmatrix} - \frac{\partial}{\partial t} \begin{bmatrix} \left( \dfrac{\partial \boldsymbol{x}}{\partial t} - \boldsymbol{f} \right) \\ \left( \dfrac{\partial \boldsymbol{\lambda}}{\partial t} + H_x \right) \\ \boldsymbol{0} \end{bmatrix} \right), \tag{65}$$

where $\boldsymbol{y} = \begin{bmatrix} \boldsymbol{x} \\ \boldsymbol{\lambda} \\ \boldsymbol{u} \end{bmatrix}$, $H = L + \boldsymbol{\lambda}^{\mathrm{T}} \boldsymbol{f}$ is the Hamiltonian, and $\boldsymbol{K}$ is a positive-definite gain matrix. By comparison, the advantages of the EPDE (33) over the AEPDE (65) may be listed as follows. i) The costates are not included and the evaluation on the second-order derivatives of $H$ is avoided, which will relieve the computation burden. ii) The solution of the EPDE is more capable of reaching a minimum, while the solution of the AEPDE may halt at a saddle. iii) Generally, the EPDE requires the integration, and the differentiation, as displayed in the AEPDE, may be avoided. This is advantageous to reduce the numerical error in seeking solutions. Regarding the disadvantages, i) currently, the EPDE may address OCPs with free terminal states only. When the terminal states are constrained, which is common in the OCP formulation, it is not directly applicable. One may penalize the terminal boundary conditions in the performance index. However, this is not a satisfactory solution. ii) Moreover, the EPDE requires the initial solution to be feasible, which is also inflexible, even if a feasible solution for the problem focused herein may be generated by the high-precision numerical integration.

The mechanism that the classic iterative methods use to solve the OCPs may be generally described as

$$(\boldsymbol{x}(t),\boldsymbol{u}(t))_{i+1}=\boldsymbol{M}\left[(\boldsymbol{x}(t),\boldsymbol{u}(t))_{i}\right], \tag{66}$$

where $\boldsymbol{M}$ represents some mapping operator and $i$ denotes the number of iteration. A vital concept introduced in the VEM is the variation time $\tau$. Interpreted from the view of the numerical computation dimension, it may be associated with the iterations in the iterative methods. Namely, the numerical iterations are the manifestation of the discrete dynamics along the virtual time dimension $\tau$. Consider in this way; the VEM proposed in this paper may be regarded as a continuous version of the iterative gradient method [33], while the AEPDE that includes the second-order Hessian derivative is a continuous realization of the Newton type iterative mechanism, in particular, with convergence theoretically guaranteed.

Regarding the computation, through the semi-discrete method, the IVPs on the infinite-dimensional dynamics may be solved with the mature ODE integration methods. Particularly during the discretization, techniques such as the Legendre-Gauss (LG), Legendre-Gauss-Radau (LGR) and Legendre-Gauss-Lobatto (LGL) discretization, which have stronger approximation capacity, may be employed to improve the accuracy and efficiency. In addition, since the integration can be achieved with a simple analog circuit, the EPDE may provide a promising way of optimal control in engineering.

# 6 Conclusions

A compact version of the variation evolving method (VEM) for the optimal control computation is developed in the primal space. This method introduces no extra variables in transforming the optimal control problems (OCPs) to the initial-value problems (IVPs). The costate-free optimality conditions are established and they are proved to be equivalent to the classic optimality conditions. One direct result found by analyzing the optimality conditions is that the analytic optimal feedback control law is generally not algebraically solvable because the control is related to the future states. Regarding the computation, with the mature ordinary differential equation (ODE) integration methods, the solution of the resulting IVP may reach the minimum of the OCP effectively. In particular, under the frame of IVPs upon continuous-time dynamics, the daunting task of searching for a reasonable step size and the annoying oscillation phenomenon around the minimum, as occurs in the discrete iterative numerical method, are eliminated. However, currently the method can solve the OCP with free terminal states only, and this restricts its application. Further studies will be carried out to address this issue.

# References

1. Pesch, H.J., Plail, M.: The maximum principle of optimal control: A history of ingenious ideas and missed opportunities. Control Cybern. 38(4), 973-995 (2009)
2. Betts, J.T.: Survey of numerical methods for trajectory optimization. J. Guid. Control Dyn. 21(2), 193-206 (1998)

3. Lin, Q., Loxton, R., Teo, K.L.: The control parameterization method for nonlinear optimal control: A survey. J. Ind. Manag. Optim. 10(1), 275-309 (2014)

4. Hargraves, C., Paris, W.: Direct trajectory optimization using nonlinear programming and collocation. J. Guid. Control Dyn. 10(4), 338-342 (1987)

5. Martin, C.S., Conway, B.A.: A new numerical optimization method based on Taylor series. In Proceedings of 21st AAS/AIAA Spaceflight Mechanics Meeting, New Orleans, 2011, AAS Paper 11-156 (2011)

6. Stryk, O.V., Bulirsch, R.: Direct and indirect methods for trajectory optimization. Ann. Oper. Res. 37(1), 357-373 (1992)

7. Peng, H.J., Gao, Q., Wu, Z.G., Zhong, W.X.: Symplectic approaches for solving two-point boundary-value problems. J. Guid. Control Dyn. 35(2), 653-658 (2012)

8. Rao, A.V.: A survey of numerical methods for optimal control. In Proceedings of AAS/AIAA Astrodynamics Specialist Conference, Pittsburgh, PA, 2009, AAS Paper 09-334 (2009)

9. Conway, B.A.: A survey of methods available for the numerical optimization of continuous dynamic systems. J. Optim. Theory Appl. 152, 271-306 (2012)

10. Bertrand, R., Epenoy, R.: New smoothing techniques for solving bang-bang optimal control problems-numerical results and statistical interpretation. Optim. Control Appl. Meth. 23(4), 171-197 (2002)

11. Pan, B.F., Lu, P., Pan, X., Ma, Y.Y.: Double-homotopy method for solving optimal control problems. J. Guid. Control Dyn. 39(8), 1706-1720 (2016)

12. Hager, W.W.: Runge-Kutta methods in optimal control and the transformed adjoint system. Numer. Math. 87, 247-282 (2000)

13. Garg, D., Patterson, M.A., Hager, W.W., Rao, A.V., et al.: A Unified framework for the numerical solution of optimal control problems using pseudospectral methods. Automatica, 46(11), 1843-1851 (2010)

14. Gong, Q., Ross, I.M., Kang, W., Fahroo, F.: Connections between the covector mapping theorem and convergence of pseudospectral methods for optimal control. Comput. Optim. Appl. 41(3), 307-335 (2008)

15. Hager, W.W., Liu, J., Mohapatra, S., Rao, A.V., Wang, X.S.: Convergence rate for a Gauss collocation method applied to constrained optimal control. SIAM J. Control Optim. 56, 1386-1411 (2018)

16. Ross, I.M., Fahroo, F.: A perspective on methods for trajectory optimization. In Proceeding of AIAA/AAS Astrodynamics Conference, Monterey, CA, 2002, AIAA Paper No. 2002-4727 (2002)

17. Fliess, M., Lévine, J., Martin, P., Rouchon, P.: Flatness and defect of nonlinear systems: Introductory theory and examples. Int. J. Contr. 61(6), 1327-1361 (1995)

18. Ross, I.M., Fahroo, F.: Pseudospectral methods for optimal motion planning of differentially flat systems. IEEE Trans. Autom. Control. 49(8), 1410-1413 (2004)

19. Milam, M.B.: Real-time optimal trajectory generation for constrained dynamical systems. Ph.D. thesis, California Institute of Technology, Pasadena, CA (2003)

20. Zhang, S., Zhao, Q., Huang, H.B., Tang, G.J.: Rapid path planning for zero propellant maneuvers based on flatness. J. Aerospace Eng. 30(5), 04017054 (2017)

21. Schropp, J.: A dynamical systems approach to constrained minimization. Numer. Math. 21, 537-551 (2000)

22. Mertikopoulos, P., Staudigl, M.: On the convergence of gradient-like flows with noisy gradient input. SIAM J. Optim. 28, 163-197 (2018)

23. Snyman, J.A., Frangos, C., Yavin, Y.: Penalty function solutions to optimal control problems with general constraints via a dynamic optimisation method. Comput. Math. Appl. 23(23), 47-55 (1992)

24. Kelley, H.J.: Methods of gradients. In: Leitmann, G.: (ed.): Optimization Techniques, pp. 216-232. Academic Press, London (1962)
25. Zhang, S., Yong, E.M., Qian, W.Q., He, K.F.: A variation evolving method for optimal control computation. J. Optim. Theory Appl. 183, 246-270 (2019)
26. Zhang, H.X., Shen, M.Y.: Computational Fluid Dynamics—Fundamentals and Applications of Finite Difference Methods. National Defense Industry Press, Beijing (2003)
27. Hartl, R.F., Sethi, S.P., Vickson, R.G.: A survey of the maximum principles for optimal control problems with state constraint. SIAM Rev. 37(2), 181-218 (1995)
28. Lou, H.W.: Analysis of the optimal relaxed control to an optimal control problem. Appl. Math. Optim. 59(1), 75-97 (2009)
29. Khalil, H.K.: Nonlinear Systems, 3rd edn. Prentice Hall, Upper Saddle River (2002)
30. Wu, D.G.: Variation Method. Higher Education Press, Beijing (1987)
31. Cassel, K.W.: Variational Methods with Applications in Science and Engineering. Cambridge University Press, Cambridge (2013)
32. Zhen, D.Z.: Linear System Theory. Tsinghua University Press, Beijing (2002)
33. Bryson, A.E., Ho, Y.C.: Applied Optimal Control: Optimization, Estimation, and Control. Hemisphere, Washington, DC (1975)
34. Wang, H., Luo, J.S., Zhu, J.M.: Advanced Mathematics. National University of Defense Technology Press, Changsha (2000)
35. Xie, X.S.: Optimal Control Theory and Application. Tsinghua University Press, Beijing (1986)
36. Patterson, M.A., Rao, A.V.: GPOPS-II: AMATLAB software for solving multiple-phase optimal control problems using *hp*-adaptive Gaussian quadrature collocation methods and sparse nonlinear programming. ACM Trans. Math. Software 41(1), 1-37 (2014)
37. Hull, D.G.: Optimal Control Theory for Applications. Springer, New York (2003)